\documentclass[12pt,preprint]{aastex}
\usepackage[version=3]{mhchem}
\usepackage{lscape}
\usepackage{longtable}

\makeatletter
\newcounter{reaction}
\renewcommand\thereaction{C\,\arabic{reaction}}
\newcommand\reactiontag{\refstepcounter{reaction}\tag{\thereaction}}
\newcommand\reaction@[2][]{\begin{equation}\ce{#2}%
\ifx\@empty#1\@empty\else\label{#1}\fi%
\reactiontag\end{equation}}
\newcommand\reaction@nonumber[1]{\begin{equation*}\ce{#1}%
\end{equation*}}
\newcommand\reaction{\@ifstar{\reaction@nonumber}{\reaction@}}
\makeatother

\shorttitle{Photochemistry-Thermochemistry in Super Earth Thick Atmospheres}

\shortauthors{Hu and Seager}

\begin{document}

\title{Photochemistry in Terrestrial Exoplanet Atmospheres III: Photochemistry and Thermochemistry in Thick Atmospheres on Super Earths and Mini Neptunes}

\author{Renyu Hu$^{1,2}$, Sara Seager$^{1,3}$}
\affil{$^1$Department of Earth, Atmospheric and Planetary Sciences, Massachusetts Institute of Technology, Cambridge, MA 02139}
\affil{$^2$Now at Division of Geological and Planetary Sciences, California Institute of Technology, Pasadena, CA 91125}
\affil{$^3$Department of Physics, Massachusetts Institute of Technology, Cambridge, MA 02139}
\email{hury@caltech.edu}

\begin{abstract}

Some super Earths and mini Neptunes will likely have thick atmospheres that are not \ce{H2}-dominated. We have developed a photochemistry-thermochemistry kinetic-transport model for exploring the compositions of thick atmospheres on super Earths and mini Neptunes, applicable for both \ce{H2}-dominated atmospheres and non-\ce{H2}-dominated atmospheres. Using this model to study thick atmospheres for wide ranges of temperatures and elemental abundances, we classify them into hydrogen-rich atmospheres, water-rich atmospheres, oxygen-rich atmospheres, and hydrocarbon-rich atmospheres. We find that carbon has to be in the form of \ce{CO2} rather than \ce{CH4} or \ce{CO} in a \ce{H2}-depleted water-dominated thick atmosphere, and that the preferred loss of light elements from an oxygen-poor carbon-rich atmosphere leads to formation of unsaturated hydrocarbons (\ce{C2H2} and \ce{C2H4}). We apply our self-consistent atmosphere models to compute spectra and diagnostic features for known transiting low-mass exoplanets GJ~1214~b, HD~97658~b, and 55~Cnc~e. For GJ~1214~b like planets we find that (1) \ce{C2H2} features at 1.0 and 1.5 $\mu$m in transmission and \ce{C2H2} and \ce{C2H4} features at 9-14 $\mu$m in thermal emission are diagnostic for hydrocarbon-rich atmospheres; (2) a detection of water-vapor features and a confirmation of nonexistence of methane features would provide sufficient evidence for a water-dominated atmosphere. In general, our simulations show that chemical stability has to be taken into account when interpreting the spectrum of a super Earth/mini Neptune. Water-dominated atmospheres only exist for carbon to oxygen ratios much lower than the solar ratio, suggesting that this kind of atmospheres could be rare.

\end{abstract}


\section{Introduction}

One of the most exciting discoveries of astronomy in recent years is the discovery of super Earths and mini Neptunes\footnote{Both super Earths and mini Neptunes correspond to the exoplanets that have masses within 10 times Earth's mass or radii within 4 times Earth's radius. Super Earths are the planets predominantly rocky in composition, whereas mini Neptunes are the planets for which a gaseous layer takes a significant fraction of the volume.} and their atmospheres. A large number of super Earths have been discovered by radial velocity surveys (e.g., Rivera et al. 2005; Udry et al. 2007; Forveille et al. 2009; Mayor et al. 2009; Vogt et al. 2010; Rivera et al. 2010; Howard et al. 2011; Bonfils et al. 2011; Dumusque et al. 2012) and transit surveys (L\'eger et al. 2009; Charbonneau et al. 2009; Holman et al. 2010; Winn et al. 2011; Demory et al. 2011; Batalha et al. 2011; Lissauer et al., 2011; Borucki et al. 2011; Cochran et al. 2011; Fressin et al 2012; Gautier et al. 2012; Borucki et al. 2012; Muirhead et al. 2012; Batalha et al. 2013; Gilliland et al. 2013; Swift et al. 2013; Dragomir et al. 2013). If a super Earth is transiting, its atmosphere may be observed via transmitted stellar radiation  (e.g., Bean et al. 2010; Croll et al. 2011; D\'esert et al. 2011; Berta et al. 2012) and planetary thermal emission (Demory et al. 2012). Detection of super Earths and characterization of their atmospheric composition with the transit technique are poised to accelerate. In the future, direct imaging will allow the super Earths to be observed in reflected light and their atmospheres to be characterized in greater details (e.g., Maire et al. 2012).

A subset of super Earths and mini Neptunes will have thick atmospheres, which we define as the atmospheres that are thick enough to maintain thermochemical equilibrium at high pressures. At high pressures, collisions between molecules in the atmosphere become so frequent that the molecular composition proceeds to thermochemical equilibrium; in other words, the Gibbs free energy of the mixture of molecules is minimized, because the system tends to reach the lowest Gibbs free energy state. Theoretical calculations show that Jupiter's atmosphere is in thermochemical equilibrium for pressures higher than $\sim1000$ bar (e.g., Fegley \& Lodders 1994), and hot Jupiter HD~189733~b's atmosphere is in thermochemical equilibrium for pressures higher than $\sim100$ bar (e.g., Moses et al. 2011). Atmospheres on super Earths/mini Neptunes, similarly to the atmospheres on Jupiter and hot Jupiters, could obtain thermochemical equilibrium at depth if the atmosphere is thick enough to reach pressures higher than $1\sim1000$ bar depending on temperature. We suggest many super Earths/mini Neptunes may have thick atmospheres because: (1) the constraints on planetary compositions from the mass and the radius of a super Earth/mini Neptune (e.g., Valencia et al. 2007b; Rogers \& Seager 2010a) usually cannot exclude the possibility of a massive gas envelope; (2) population synthesis studies of planet formation have suggested planets that are more massive than Earth may accrete much more volatiles than Earth and are more likely to have thick atmospheres (e.g., Mordasini et al. 2012; Fortney et al. 2013).

Some super Earths and mini Neptunes may have thick atmospheres that are not mainly composed of hydrogen. Super Earths and mini Neptunes obtain their atmospheres by capture from the nebula, degassing during accretion, and degassing from tectonic processes (Elkins-Tanton \& Seager 2008). First, starting out as an \ce{H2}-dominated atmosphere accreted from the planet-forming nebula, the atmosphere on a super Earth/mini Neptune can evolve to become a non-\ce{H2}-dominated atmosphere. This is because super Earths and mini Neptunes may have experienced more significant atmospheric loss than gas giants because they are not as massive as gas giants (Yelle 2004; Yelle et al. 2008; Lammer et al. 2008, 2013). The result of atmospheric evolution on some super Earths and mini Neptunes should be preferential loss of light atoms (i.e., H and He) and enrichment of more heavy atoms (e.g., C, O, N, S). Second, a thick atmosphere may be formed by degassing during accretion on a super Earth/mini Neptune and its composition may be non-\ce{H2}-dominated. Although the outcome of accretion degassing depends on the accretion process and the material being accreted, it can be generally expected that a few percent of the planet's mass can be degassed to form an extensive atmosphere composed of hydrogen, oxygen, and carbon compounds (Elkins-Tanton \& Seager 2008). Recent observations of GJ~1214~b, a super Earth/mini Neptune of $\sim6$ M$_{\earth}$, have hinted that the planet has an atmosphere with a mean molecular mass significantly larger than that of a hydrogen-dominated atmosphere (e.g, Bean et al. 2010; Berta et al. 2012). 

This work focuses on thick atmospheres of hydrogen, carbon, and oxygen as they are likely to be the most important building blocks of super Earth atmospheres, and their relative abundance controls the molecular composition in thermochemical equilibrium. We are motivated to consider a wide range of carbon to oxygen elemental abundance ratio (i.e., the C/O ratio) for thick atmospheres and here is why. First, if the thick atmosphere is formed by capturing the gases and ices in the planet-forming nebula, its C/O ratio should to a certain extent track the C/O ratio of the nebula and therefore the parent star (Bond et al. 2010; Madhusudhan et al. 2012). Though still debated (Fortney 2012; Nissen 2013), stellar observations have shown that the C/O ratio of planet-hosting systems spreads over a wide range between 0.3 and 2.0 (Bond et al. 2010; Delgado Mena et al. 2010; Petigura \& Marcy 2011; Teske et al. 2013). A significant fraction of planet-hosting systems may have C/O ratios greater than the solar C/O ratio (Petigura \& Marcy 2011; Fortney 2012). Protoplanetary nebulae that are enriched in carbon have a different condensation sequence than the Solar System, which may lead to the formation of carbon-rich terrestrial planets, i.e. carbon planets (Lodders 2004; Kuchner \& Seager 2005; Bond et al. 2010; Johnson et al. 2012). 
%Indeed, an atmosphere that has more carbon than oxygen has been suggested for a hot Jupiter (WASP-12b, Madhusudhan et al. 2011). 
%For carbon planets that have carbon-rich atmospheres (but still \ce{H2}-dominated), a number of studies have anticipated high abundance of \ce{CO} and scarcity of \ce{H2O} (Kuchner \& Seager 2005; Madhusudhan et al. 2011b; Kopparapu et al. 2012; Moses et al. 2013). 
Second, for a thick atmosphere originated from degassing of accretion of undifferentiated chondritic materials, the composition can be either oxygen-rich or carbon-rich depending on the nature of accreted materials. To be specific, accretion of group CI and CM carbonaceous chondrites would produce \ce{H2O}-dominated atmospheres, accretion of group CR, CO, and CH carbonaceous chondrites would produce highly carbon-rich atmospheres with the C/O ratio greater than 1, and outgassing of ordinary and enstatite chondrites would produce atmospheres with the C/O ratio close to 1 and various amounts of \ce{H2} (Elkins-Tanton \& Seager 2008; Schaefer \& Fegley 2010).
After all, due to the uncertainty of the formation and evolution processes of planetary atmospheres, the C/O ratio of the protoplanetary nebula does not necessarily align with that of the planet formed therein, nor the C/O ratio of the planetary atmosphere necessarily aligns with that of the bulk planet mass. It is therefore plausible to consider super Earth atmospheres to have a wide range of C/O ratios, from much lower than the solar ratio (0.5) to much higher than unity.

A model for thick atmospheres should compute chemical reaction kinetics and vertical transport in three pressure regimes: the low-pressure regime where photon-driven processes dominate; the intermediate regime where vertical transport dominates; and the high-pressure regime where thermochemical equilibrium dominates. We call such a model ``photochemistry-thermochemistry kinetic-transport model'' or simply ``photochemistry-thermochemistry model''\footnote{Such models is sometimes called ``photochemistry models" in the context of discriminating from thermochemistry models. However, this might cause confusion, because the term ``photochemistry model" usually implies photochemical processes driven by parent-star irradiation are the dominant factor for molecular compositions in the modeled atmospheres.}.

A photochemistry-thermochemistry model is critical for the study of molecular compositions of any thick atmospheres, including the thick atmospheres on super Earths and mini Neptunes. The overarching reason is that the composition of the observable part of a thick atmosphere (0.1 mbar  to 1 bar, depending on the wavelength) is controlled by both chemical reactions and vertical transport.  For thick atmospheres there is a competition between chemical reactions that drive the system to thermochemical equilibrium and vertical transport that tends to mix the atmosphere bringing parts of the atmosphere to disequilibrium. The division between the thermochemical equilibrium regime and the transport-driven disequilibrium regime is the so called ``quenching pressure''. The quenching pressure has been worked out for \ce{H2}-dominated thick atmospheres on Jupiter and Saturn (Prinn \& Barshay 1977; Prinn \& Olaguer 1981; Fegley \& Prinn 1985; Fegley \& Lodders 1994), and some brown dwarfs and hot Jupiters (Fegley \& Lodders 1996; Griffith \& Yelle 1999; Cooper \& Showman 2006; Line et al. 2010; Visscher \& Moses 2011; Moses et al. 2011). The key point is that 
%although ultraviolet photons usually {\bf cannot affect the major components of the atmosphere} above 1 mbar, 
a full photochemistry-thermochemistry model that treats chemical kinetics and vertical transport is required to compute the quenching pressure and the molecular composition in the observable part of the atmosphere.

For completeness we summarize previous work on photochemistry-thermochemistry models for exoplanet atmospheres. Models of extrasolar gas giants with thick atmospheres (by definition \ce{H2}-dominated and with codes applicable to the pressure levels at which thermochemical equilibrium is reached) have been developed (Liang et al. 2003; Zahnle et al. 2009a,b; Line et al. 2010; Visscher \& Moses 2011; Moses et al. 2011; Kopparapu et al. 2012). Recently, models of super Earth/mini Neptune GJ~1214~b having \ce{H2}-dominated thick atmospheres have been developed by Miller-Ricci Kempton (2012). Terrestrial exoplanets with thin atmospheres (such that achieving thermochemical equilibrium at the surface is kinetically prohibited) having a wide range of chemical composition have been studied by Selsis et al. (2002), Segura et al. (2005, 2007), Domagal-Goldman et al. (2011), Hu et al. (2012, 2013), and Rugheimer et al. (2013).

The super Earths and mini Neptunes being observed represent a new situation of atmospheric chemistry in the middle between extrasolar giant planets and Earth-like exoplanets, for which a self-consistent photochemistry-thermochemistry model applicable to non-\ce{H2}-dominated atmospheres has not yet been developed. When the atmosphere is not \ce{H2}-dominated (but could still contain hydrogen), the atmospheric chemistry, especially the transport quenching of key species, will be different from the case of \ce{H2}-dominated atmospheres. A unique challenge in developing a generic photochemistry-thermochemistry model for non-\ce{H2}-dominated thick atmospheres is that the model cannot assume a fixed \ce{H2} background atmosphere (taken for granted in previous work for atmospheres on gas giants). We aim at developing a photochemistry-thermochemistry model that can compute not only the trace gases, but also the abundances of major gases in thick atmospheres. 

% Another challenge in developing such a model is to properly determine the lower boundary conditions. Per definition of thick atmospheres, the lower boundary can be chosen as the pressure level at which thermochemical equilibrium of all species is achieved. In principle, disequilibrium chemistry can extend to higher pressures for lower temperatures and more effective vertical mixing. In our model, we assume a constant-abundance lower boundary condition according to thermochemical equilibrium, which depends on the temperature at the lower boundary. The temperature, in turn, depends on the atmospheric composition. Hence, the lower boundary conditions have to be derived iteratively from a coupled thermochemistry code and radiative transfer code before launching the photochemistry simulation. 
%Moses et al. (2011) assumed zero-flux boundary conditions, i.e., no mass can be transported into or out from the system. Such an assumption inevitably introduces dependency on the initial conditions, as a mass-conserved model would have to carry on the initial conditions to the results. However, we found in our simulations the initial conditions shall have no effects on the steady-state composition if the boundary conditions are properly chosen. 

In this work we present the first comprehensive photochemistry-thermochemistry model for non-\ce{H2}-dominated thick atmospheres on super Earths and mini Neptunes. We present simulations of exoplanet thick atmospheres with thermochemical equilibrium maintained at depth and photochemical processes and vertical transport controlling the compositions in the observable part of the atmospheres. Section~2 describes our photochemistry-thermochemistry model, including the fundamental parameters for the model and numerical tests for model validation. We present our main findings regarding the molecular compositions of thick atmospheres on super Earths and mini Neptunes in Section~3. We apply our self-consistent atmosphere models to known transiting low-mass exoplanet GJ~1214~b, HD~97658~b, and 55~Cnc~e, and present their synthetic spectra based on the photochemistry-thermochemistry simulations to provide observable predictions in Section~4. Section~5 discusses the formation conditions for water-dominated atmospheres, chemical stability of gases in thick atmospheres, and the effects of disequilibrium chemistry in thick atmospheres. We conclude in Section~6.

\section{Model}

\subsection{Photochemistry-Thermochemistry Model}

\begin{figure}[h]
\begin{center}
 \includegraphics[width=1.0\textwidth]{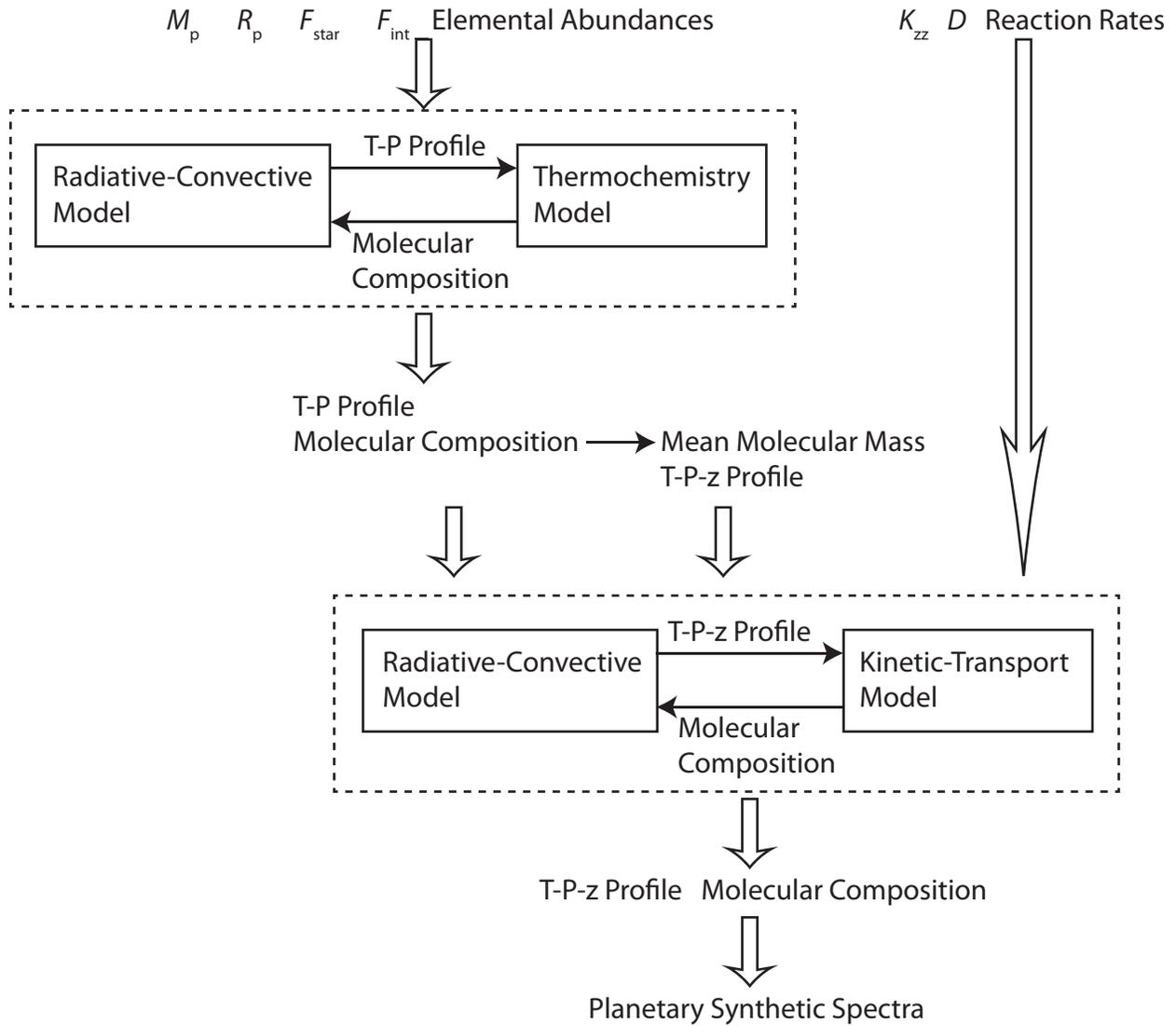}
 \caption{
 Architecture of the photochemistry-thermochemistry kinetic-transport model for thick atmospheres on exoplanets.
 $M_{\rm p}$ and $R_{\rm p}$ are the mass and radius of the modeled planet, $F_{\rm star}$ and $F_{\rm int}$ are the stellar and internal heat fluxes that control the temperature of the planet's atmosphere, and $K_{\rm zz}$ and $D$ are the vertical diffusion coefficients for atmospheric eddy and molecular diffusion.
  }
 \label{ModelSchematic}
  \end{center}
\end{figure}

Figure \ref{ModelSchematic} schematically show the architecture of our photochemistry-thermochemistry kinetic-transport model for thick atmospheres on exoplanets. We design our model to have two levels of sophistication: the first level is a thermochemical equilibrium model to compute the atmospheric composition at thermochemical equilibrium, with the temperature-pressure profile computed by a coupled radiative-convective model; the second level, beyond the thermochemical equilibrium model, is a kinetic-transport model that treats the effects of vertical mixing and photochemical processes on the molecular composition, with the temperature-pressure profile self-consistently computed based on the molecular composition from disequilibrium chemistry. The first-level thermochemistry model provides appropriate initial conditions, including the atmosphere's thermal structure and molecular composition, for the second-level kinetic-transport model.

A unique feature in our photochemistry-thermochemistry model for non-\ce{H2}-dominated thick atmospheres is that the model does not require specification of the main component of the atmosphere (nor the mean molecular mass) and instead the model takes the elemental abundances as the input parameters. All previous photochemistry models assume a specific dominant gas (and therefore mean molecular mass) and seek steady-state abundances of trace gases in the fixed background atmosphere. However, for applications to super Earths, one cannot assume any specific dominant gas, and the mean molecular mass needs to be self-consistently determined. We provide for the first time such a feature in our photochemistry-thermochemistry code. In the first-level thermochemistry model that includes a thermochemical equilibrium routine and a radiative-convective routine, we use a pressure level grid so that the mean molecular mass is no longer required. The mean molecular mass is synthesized from the thermochemical equilibrium composition profile, and then used to transform the pressure grid to an altitude grid. In the second-level photochemistry-thermochemistry model, we perform kinetic-transport simulation on a pressure-temperature-altitude profile which itself is updated by the radiative-convective calculation. Our approach eliminates the need to specify a background atmosphere for kinetic-transport simulations of thick atmospheres, which makes our model uniquely suitable for applications in the study of super Earths and mini Neptunes.

\subsubsection{Thermochemisty Model}

A thermochemical equilibrium model and a radiative transfer model are used iteratively to obtain stratified atmospheric compositions and temperature-pressure profiles that obey thermochemical equilibrium, radiative-convective equilibrium, and hydrostatic equilibrium. We use the method of minimizing the total Gibbs free energy as described in Miller-Ricci et al. (2009) to compute the thermochemical equilibrium composition based on elemental abundances. The temperature profiles are computed based on atmospheric compositions that obey thermochemical equilibrium. We compute the temperature-pressure profiles by balancing the stellar irradiation with planetary thermal emission. With a grey atmosphere assumption (i.e., mean opacities in stellar radiation wavelengths and in thermal emission wavelengths), we compute the temperature-pressure profiles using the formulation of Guillot (2010). The opacities include absorption cross sections of \ce{CO2}, \ce{CO}, \ce{CH4}, \ce{H2O}, \ce{O2}, \ce{O3}, \ce{OH}, \ce{CH2O}, \ce{CH2O2}, \ce{H2O2}, \ce{HO2}, \ce{C2H2}, \ce{C2H4}, and \ce{C2H6} generated from the HITRAN 2008 database (Rothman et al. 2009), and \ce{H2}-\ce{H2} collision-induced absorption from Borysow (2002). The temperature-pressure profiles are adjusted according to the appropriate adiabatic lapse rate to account for the onset of convection (see Miller-Ricci et al. 2009). When the thermochemical equilibrium model and the radiative-convective model converge, we calculate the mean molecular mass of the atmosphere as a function of pressure and project the temperature-pressure profile to an altitude grid.

\subsubsection{Kinetic-Transport Model}

The kinetic-transport model solves the one-dimensional continuity-transport equation for the steady state, viz.
\begin{equation}
\frac{\partial n}{\partial t} = P - L - \frac{\partial \Phi}{\partial z}, \label{EqF}
\end{equation}
where $n$ is the number density for a certain species (cm$^{-3}$), $z$ is the altitude, $P$ and $L$ are the production and loss rates of the species (cm$^{-3}$ s$^{-1}$), and $\Phi$ is the vertical transport flux of the species (cm$^{-2}$ s$^{-1}$). The production and loss rates are the summary of all contributions from relevant chemical and photochemical reactions in the atmosphere. The transport flux that couples different layers of the atmosphere is parameterized by eddy diffusion and molecular diffusion, i.e.,
\begin{equation}
\Phi=-K_{\rm zz}N\frac{\partial f}{\partial z} - DN\frac{\partial f}{\partial z}
+Dn\bigg(\frac{1}{H_0}-\frac{1}{H}-\frac{\alpha_{\rm T}}{T}\frac{dT}{dz}\bigg) ,\label{eq_diffusion}
\end{equation}
where $K_{\rm zz}$ is the eddy diffusion coefficient (cm$^2$ s$^{-1}$), $D$ is the molecular diffusion coefficient (cm$^2$ s$^{-1}$), $N$ is the total number density of the atmosphere, $f\equiv n/N$ is the mixing ratio of the species, $H_0$ is the mean scale height, $H$ is the scale height of the species, $T$ is the temperature (K), and $\alpha_{\rm T}$ is the thermal diffusion factor. The first term of Equation (\ref{eq_diffusion}) represents eddy diffusion, and the last two terms represent molecular diffusion. Equations (\ref{EqF} and \ref{eq_diffusion}) are solved numerically by the inverse-Euler method with the lower boundary condition set by thermochemical equilibrium.

Physical processes presented by $P$, $L$, and $\Phi$ in Equation (\ref{EqF}) are competing and which one dominates depends on the temperature and pressure, i.e., altitude. Deep in the atmosphere, $P$ and $L$ are the dominant terms, because they increase with pressure and temperature, and the steady-state condition becomes $P=L$, i.e., thermochemical equilibrium. Near the top of the atmosphere, photochemical processes contribute dominantly to the $P$ and $L$ terms. The effect of photochemical processes confines to pressure levels above 0.1 bar, because ultraviolet photons that could dissociate atmospheric molecules usually only penetrate to the pressures of 0.1 bar due to Rayleigh scattering and absorption by molecules in the upper atmosphere. In the intermediate pressure levels, the transport flux $\Phi$ can be the major source and sink for species as compared with photochemical and chemical reactions. The reason is that the reactive radicals produced by photochemistry are usually not abundant because of the lack of photons that penetrate to these altitudes, and that the temperature and pressure (and therefore reaction rates) are usually not high in the intermediate pressures. In all, when all production and loss mechanisms are properly taken into account in $P$, $L$ and $\Phi$,  Equation (\ref{EqF}) is the governing equation for molecular compositions in any atmospheres.

We have extended the photochemistry model described in Hu et al. (2012) to the pressure regime in which thermochemical equilibrium holds. To make the model applicable to the pressure regime of thermochemical equilibrium, we include the contribution of the reverse equation for each forward equation into the production and loss terms in Equation (\ref{EqF}). The reverse reactions are the reversal of the chemical reactions whose kinetic rates have been measured or computed for the studies of Earth and planetary atmospheres (chosen as forward reactions). The kinetic rates of reverse reactions ($k_r$) and forward reactions ($k_f$) are linked by the difference in the Gibbs free energy of formation ($\Delta_fG^{\circ}$) of the reactants and the products as
\begin{equation}
k_r = k_f \exp{\bigg[\frac{\Delta_fG^{\circ}({\rm products})-\Delta_fG^{\circ}({\rm reactants})}{RT}\bigg]} (k_b'T)^{n_p-n_r}, \label{reverse}
\end{equation}
where $R$ is the gas constant, $T$ is the temperature (K), $k_b'=1.38065\times10^{-22}$ bar cm$^3$ K$^{-1}$ is the Boltzmann constant, and $n_p$ and $n_r$ are the number of products and reactants, respectively (Visscher \& Moses 2011).

Using Equation (\ref{reverse}), we calculate the rates of the reverse reaction of each forward reaction in the reaction list of Hu et al. (2012) that includes bimolecular reactions, termolecular reactions, and thermodissociation reactions. The Gibbs free energies of formation are taken from the NIST-JANAF database (Chase 1998), Burcat \& Ruscic (2005), and Visscher \& Moses (2011). For a number of reactions, there are empirical measurement for both forward and reverse reactions, which provide an opportunity of validating this computation scheme. Whenever possible, we adopt the empirical kinetic rate that have the widest temperature range (i.e., define as the forward reaction), compute the reverse reaction rate, and compare the reverse rate with its empirical measurements. For all tested reactions we find agreement within one order of magnitude, which validates our calculation of reverse reaction rates. 

Another update to our reaction network for the model to be applicable to high pressures is that we have included the high-pressure limit rates for all three-body reactions and dissociation reactions. In the reaction list of Hu et al. (2012) many three-body reactions only have low-pressure limit rates, which are appropriate for applications to thin atmospheres. We have added the high-pressure limit rates for the three-body reactions and thermo-dissociation reactions, based on recommended rates of the NIST Kinetics database\footnote{http://kinetics.nist.gov/}, Baulch et al. (1994), Baulch et al. (2005), Jasper et al. (2007), and Moses et al. (2011).
 
%For simplicities we do not update the temperature-pressure profile in the kinetic-transport simulations, although the composition profile may change. One might wonder how temperatures would change as a result of vertical transport and photochemical processes. We have verified that these processes only result in minor alteration in the temperatures, usually under xx K in terms of the temperature in the bottom layer. This justifies our simplification.}

%The temperature-pressure profile and the composition at the lower boundary are then provided to the full kinetic-transport model to seek the steady-state solution.

We have coupled our kinetic-transport model to the grey-atmosphere radiative-convective model. As the composition (and therefore the mean molecular mass) might change as the photochemistry code steps forward, the temperature-pressure profile and the pressure-altitude conversion need to be updated to keep the atmosphere in radiative equilibrium and hydrostatic equilibrium. We therefore iteratively run the kinetic-transport routine and the radiative-convective routine for this update. In practice, the altitude grid is not re-calculated for each time step in kinetic-transport simulations. Instead, as chemistry only has secondary dependency on the atmospheric temperature, we run the kinetic-transport code to convergence, and use the radiative-convective code to calculate the a temperature-pressure-altitude profile based on the converged kinetic-transport result. Then, the new atmospheric profile, with the new mean molecular mass incorporated, is provided to the kinetic-transport code for another run. Starting from the atmospheric structure profile provided by the first-level thermochemistry model, we find that the temperature profile typically converges with the disequilibrium chemistry composition within 3 iterations. Here, by convergence we practically mean the variation of temperature between two successive iterations to be within 2\%. The end results of atmospheric composition and temperature profile satisfy radiative-convective equilibrium, hydrostatic equilibrium, and the detail balance between production and loss by chemical and photochemical processes and vertical transport.

For the current model we focus on thick atmospheres on warm and hot super Earths/mini Neptunes, in which water vapor will not condense. Table \ref{critical} tabulates the critical temperatures of common gases in planetary atmospheres. Water vapor, having a critical temperature of 647 K, stands out as having a much higher critical temperature than any other common volatiles made of C, H, O, and N elements. Therefore it is possible for a certain class of planets that have equilibrium temperatures lower than 647 K, water vapor could condense to form oceans, and then the planets could be potentially habitable.
%(see Rogers \& Seager 2014 for a detailed analysis on the parameter regime in which liquid water could be present on super Earths and mini Neptunes) 
For this paper, we focus on the atmospheres in which water remains in its gas phase. We emphasize that most of the transiting super Earths and mini Neptunes discovered so far fall within our consideration. Note that some minor constituents of the atmosphere, not included in our photochemistry-thermochemistry model, might condense out to form layers of cloud or haze. In fact, for super Earth/mini Neptune GJ~1214~b and perhaps other loss-mass exoplanets potential layers of condensates could mask molecular features in their transmission spectra and affect interpretation of the spectra (e.g., Fortney 2005; Benneke \& Seager 2012; Morley et al., 2013).

\subsubsection{Synthetic Spectrum Model}

After the photochemistry-thermochemistry simulation converges to a steady state, we compute the synthetic spectra of the modeled exoplanet's atmospheric transmission, reflection and thermal emission with a line-by-line radiative transfer code developed based on the concept outlined by Seager \& Sasselov (2000) and Seager et al. (2000). Opacities are based on molecular absorption with cross sections computed from the HITRAN 2008 database (Rothman et al. 2009), molecular collision-induced absorption when necessary (Borysow 2002), Rayleigh scattering, and aerosol extinction computed based on the Mie theory (Van de Hulst 1981). Empirical absorption coefficients of \ce{CH4} at 750-940 nm are also included (O'Brien \& Cao 2002). The transmission is computed for each wavelength by integrating the optical depth along the limb path, as outlined in Seager \& Sasselov (2000). The reflected stellar light and the planetary thermal emission are computed by the $\delta$-Eddington 2-stream method (Toon et al. 1989).

\subsection{Fundamental Parameters for the Photochemistry-Thermochemistry Model}

For a super Earth/mini Neptune with known stellar irradiation level, the controlling factors for the molecular composition of its thick atmosphere are the elemental abundance, the efficiency of vertical mixing, and the internal heat flux. The elemental abundance determines the molecular composition at thermochemical equilibrium, the efficiency of vertical mixing determines to which extent vertical transport brings up gases from the thermochemical equilibrium regime, and internal heat flux determines the temperatures at high pressures that affect the compositions at thermochemical equilibrium. The elemental abundance rather than the background molecular composition is a fundamental parameter in our models for thick atmospheres, because an arbitrary combination of molecules may not be chemically stable in thick atmospheres (a point that will be discussed later in Section~5).

We emphasize the uniqueness of the photochemistry-thermochemistry solution. In mathematical terms, the mixing ratio profiles at the steady state obeys the continuity equation, a second-order ordinary differential equation whose solution is uniquely determined by the boundary conditions at both ends. For the upper boundary conditions, we place a zero-flux lid at the pressure level of $10^{-3}$ Pa for all gases except \ce{H} and \ce{H2}, for which we apply the diffusion-limited escape rates. As such, our solutions of atmospheric chemical composition, for a certain planetary scenario with an assigned value for the eddy diffusion coefficient, is uniquely determined by the lower boundary conditions, for which we apply the thermochemistry equilibrium compositions. We have run a number of test cases to verify the uniqueness from model simulations. For example, for an H:O:C ratio of 2:2:1, the molecular composition can be either 50\% \ce{H2} 50\% \ce{CO2}, or 50\% \ce{H2O} 50\% \ce{CO}. Under thermochemical equilibrium, the deep atmosphere will be composed of 50\% \ce{H2} and 50\% \ce{CO2}; in a separate simulation, we start with 50\% \ce{H2O} 50\% and \ce{CO} as the initial condition, the system naturally evolves to the steady-state composition of 50\% \ce{H2} 50\% \ce{CO2} at high pressures, i.e., the unique solution. The photochemistry-thermochemistry solution of the atmospheric chemical composition on an exoplanet is the only steady-state solution for a certain elemental abundance, vertical mixing rate, and internal heat flux.

\subsection{Model Validation}

\begin{figure*}[h]
\begin{center}
 \includegraphics[width=1.0\textwidth]{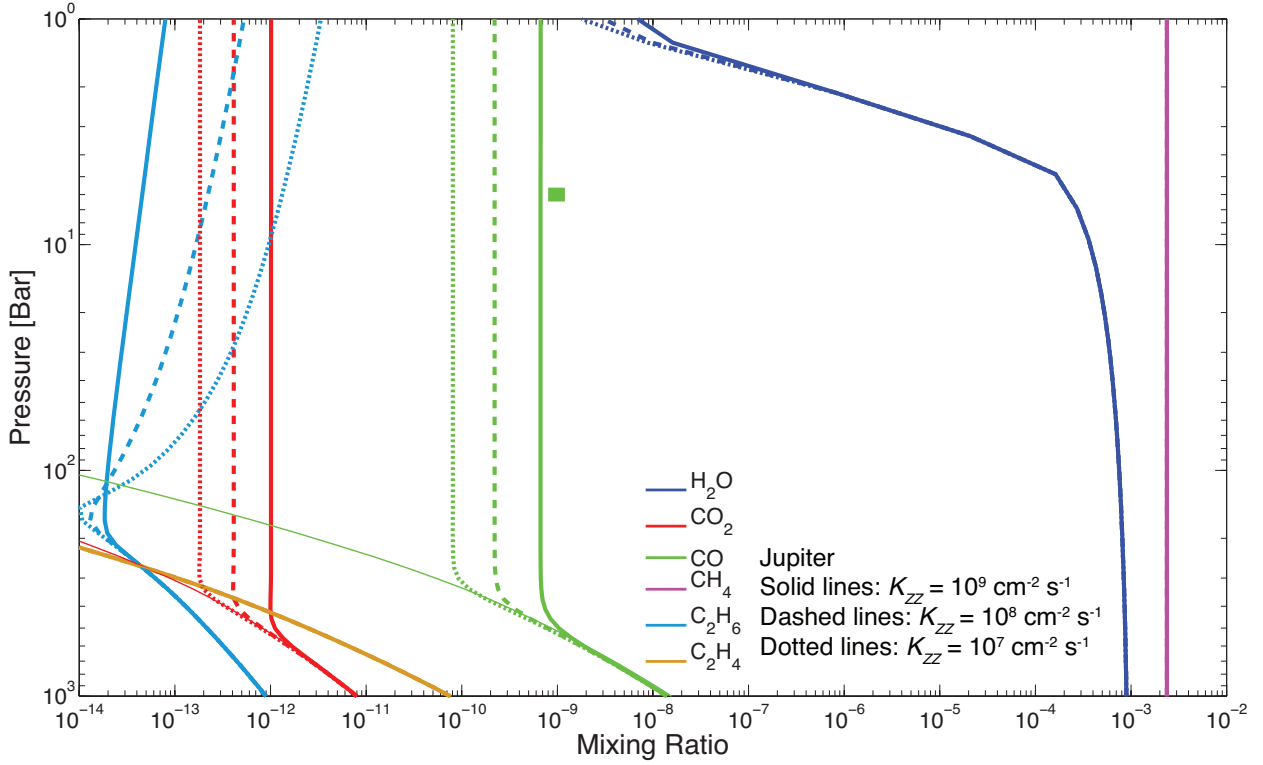}
 \caption{
 Modeled compositions of the deep atmosphere of Jupiter in comparison with observations. The mixing ratios of \ce{H2O}, \ce{CH4}, \ce{CO}, \ce{CO2}, \ce{C2H6}, and \ce{C2H4} are shown as a function of the atmospheric pressure, for eddy diffusion coefficients ranging from $10^7$ to $10^9$ cm$^2$ s$^{-1}$ according to free-convection and mixing-length theories (Gierasch \& Conrath 1985; Visscher et al. 2010), and the mixing ratios at thermochemical equilibrium are shown by thin solid lines for comparison.
 The temperature profile is adopted from Galileo probe measurements and Cassini CIRS measurements (Seiff et al. 1998; Simon-Miller et al. 2006). 
 The abundance of \ce{CH4} is assigned to be consistent with measurements of the Galileo probe at pressure levels deeper than 10 bar (Wong et al. 2004), and the abundance of \ce{H2O}  at the bottom is assigned to be the solar abundance.
 The computed mixing ratios of \ce{CO} are compared with the measurements by high-resolution spectroscopy at 6 bar (B\'ezard et al. 2002; thick horizontal bar on the figure). 
Our model is able to predict reasonably well the degree of enhancement of \ce{CO} in Jupiter's atmosphere due to vertical transport.
  }
 \label{Jupiter}
  \end{center}
\end{figure*}

\begin{figure*}[h]
\begin{center}
 \includegraphics[width=1.0\textwidth]{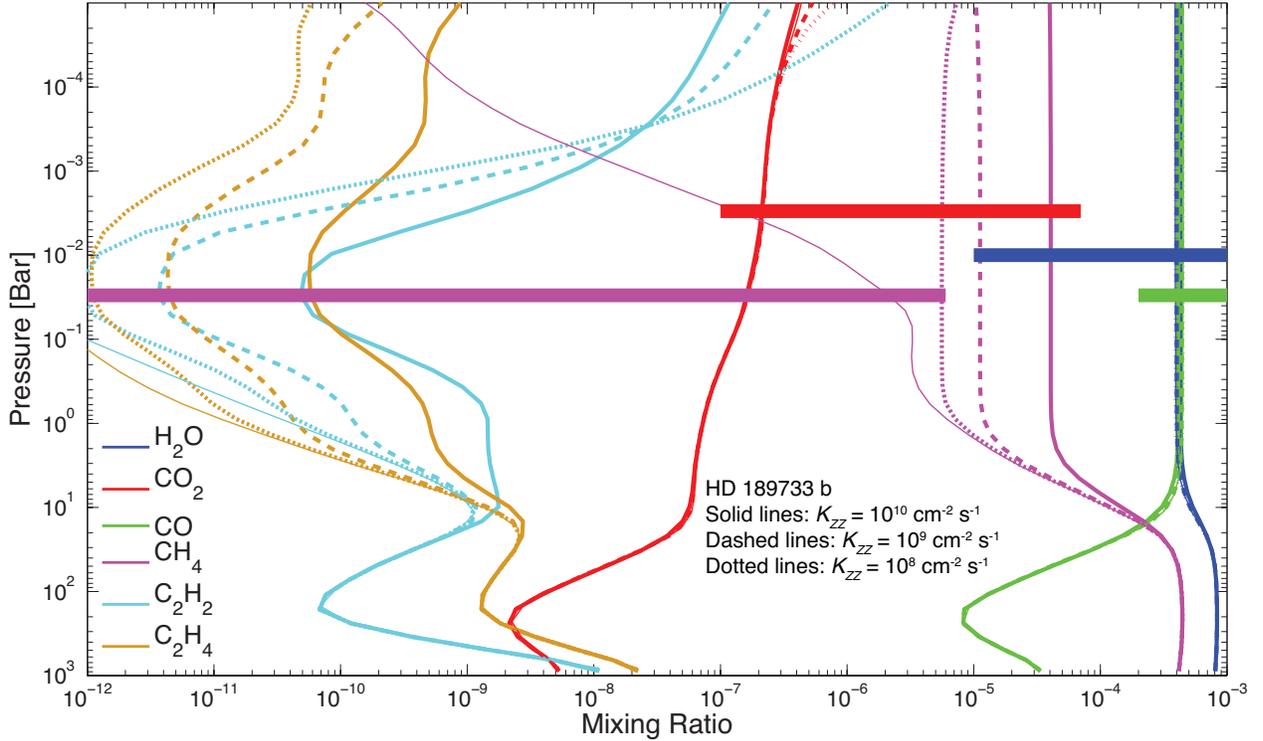}
 \caption{
 Mixing ratios of common molecules in simulated atmospheres of hot Jupiter HD~189733~b. The mixing ratios of \ce{H2O}, \ce{CH4}, \ce{CO}, \ce{CO2}, \ce{C2H2}, and \ce{C2H4} are shown as a function of the atmospheric pressure for three eddy diffusion coefficients, and the mixing ratios at thermochemical equilibrium are shown by thin solid lines for comparison. The temperature profiles are the adopted dayside averaged temperature profiles in Moses et al. (2011). The mixing ratios of \ce{H2O}, \ce{CH4}, \ce{CO}, and \ce{CO2} inferred from {\it HST} and {\it Spitzer} observations by Madhusudhan \& Seager (2009) are shown with thick horizontal bars for comparison. The pressure levels at which these bars are placed are arbitrarily chosen within $10^{-3}$ - 1 bar for illustration. Our simulations agree with the interpretation of observations for HD~189733~b, and the upper limit of \ce{CH4} mixing ratio could places an upper limit on the eddy diffusion coefficient of the planet.
  }
 \label{HD189}
  \end{center}
\end{figure*}

We have validated our photochemistry-thermochemistry model in pressure regimes for different physical processes. The model is applicable to the low pressures where photochemistry dominates, the intermediate pressures where vertical transport dominates, and the high pressures where thermochemical equilibrium dominates. For the low pressures where photochemistry dominates, the model in this paper reduces to the photochemistry model for thin atmospheres, which has been validated in Hu et al. (2012). For the high pressures where thermochemical equilibrium dominates, we find that our kinetic model that balances all forward and reverse reactions gives identical results compared with direct minimization of the global Gibbs free energy (Figure \ref{Jupiter} and Figure \ref{HD189}). This straightforward mathematical validation implies that our calculation of reverse reaction rates is correct and our model is applicable for the regime of thermochemical equilibrium. 

For the intermediate pressures in which vertical transport dominates, direct mathematical validation is not possible, because the steady-state solution is unknown. Instead, we seek to validate our model in this regime by reproducing the results of observations of planets and other models. We have simulated the atmosphere of Jupiter and the atmosphere of hot Jupiter HD~189733~b, to compare with observations. We focus on reproducing the effect of vertical transport on the \ce{CO <=> CH4} conversion in these atmospheres that have very different temperatures (Figure \ref{Jupiter} and Figure \ref{HD189}). We describe the details of these model tests below.

%We have tested our photochemistry-thermochemistry model by simulating the C-H-O chemistry in the deep atmosphere of Jupiter. 
For the deep atmosphere of Jupiter, the most important feature relevant for this investigation is the enhancement of \ce{CO} mixing ratio in the observable parts of the atmosphere due to vertical transport. Our model correctly predicts this enhancement in comparison with the observed mixing ratio of \ce{CO} (Figure \ref{Jupiter}). We have also compared our simulated \ce{CO} mixing ratios with those of the latest model for Jupiter's deep atmosphere (Visscher et al. 2010) and found agreement within a factor of 2. The remaining discrepancy may be due to updated kinetic rates of several reactions that we have adopted. Compared with reaction rates used in Visscher et al. (2010) and Moses et al. (2011), we found that the rates for reaction \ce{CH3O -> CH2O  + H} have updated recommended values and the discrepancy in these rates can account for the minor discrepancy in the \ce{CO <=> CH4} conversion rates between our models and the models of Visscher et al. (2010). We have taken for this reaction the low-pressure limit rate from Baulch et al. (1994) and the high-pressure limit rate from Curran (2006); both recommended values are based on extensive literature reviews.

To test the \ce{CO <=> CH4} conversion computation for a warmer planet than Jupiter, we have simulated the \ce{H2}-dominated atmosphere on hot Jupiter HD~189733~b and compared our predictions with the molecular compositions derived from observations (Figure \ref{HD189}). We find agreement in the amounts of \ce{CH4} and \ce{CO2} at the pressure levels of 0.001-1 bar between our models and the interpretation of observations of Madhusudhan \& Seager (2009). Due to limited data available for exoplanets including HD~189733~b, the observational constraints on the molecular composition are poor. To further test our model, we have compared our results with a similar suite of simulations by Visscher \& Moses (2011) and found agreement to within a factor of 2. The efficiency of vertical mixing affects the steady-state mixing ratios of methane and other hydrocarbons significantly in the observable parts of the atmosphere. In particular, the eddy diffusion coefficient needs to be no greater than $10^8$ cm$^2$ s$^{-1}$ to produce a methane mixing ratio consistent with the observationally derived upper limit.

Furthermore, we have also compared our results with the main features of \ce{H2}-dominated atmospheres simulated by Miller-Ricci Kempton et al. (2012) for super Earth/mini Neptune GJ~1214~b. For a solar elemental abundance, we simulate the steady-state atmosphere on GJ~1214~b with C, H, O, N chemistry. Our simulations compare nicely with Miller-Ricci Kempton et al. (2012) in both qualitative and quantitative behaviors of \ce{H2O}, \ce{CH4}, \ce{NH3} and \ce{N2}. We see also photochemical \ce{HCN} formation in the upper part of atmosphere, similar to Miller-Ricci Kempton et al. (2012); however, we find 1-order-of-magnitude higher concentration of \ce{HCN} and 1-order-of-magnitude lower concentration of \ce{C2H2} and \ce{C2H4} than Miller-Ricci Kempton et al. (2012). We suspect that this is because we have included a more complete set of chemical and photochemical reactions that allows the formation of C-N bond as Moses et al. (2011), and then a larger fraction of \ce{CH3} from methane photolysis is converted into \ce{HCN} instead of forming \ce{C2H2} and \ce{C2H4}. In summary, we find our model is able to simulate \ce{H2}-dominated thick atmosphere of irradiated exoplanets; and we here extend the analysis to irradiated thick atmospheres that are not \ce{H2}-dominated.

\subsection{Model Atmospheres in This Paper}

With the photochemistry-thermochemistry model, we simulate the molecular composition of thick atmospheres on super Earths and mini Neptunes for wide ranges of temperature and elemental abundance. We focus mostly on exploring different temperatures and elemental abundances as they are the most important parameters for the bulk compositions of the thick atmospheres. Furthermore we also explore the effect of stellar spectrum and eddy diffusivity. For each scenario, we use the thermochemistry-photochemistry model to compute the steady-state molecular composition from $10^4$ bar to $10^{-8}$ bar. We verify that the lower boundary at $10^4$ bar is sufficient to maintain thermochemistry equilibrium at the lowest atmosphere layer for each simulation. Typically, we find suitable to assign \ce{O(^1D)}, \ce{C}, and \ce{^1CH2} to be ``fast species" in the simulations for \ce{H2}-dominated cases, and \ce{O(^1D)}, \ce{^1CH2}, and \ce{CH2O2} to be ``fast species" in the simulations for non-\ce{H2}-dominated cases. The specifics and the rationale for exploring the relevant parameters are outlined in the following.

{\it Elemental Abundances of the Atmosphere} With the photochemistry-thermochemistry model, we simulate thick atmospheres of exoplanets with a variety of C-H-O elemental abundances. We focus on the C-H-O chemistry as they are the most common elements in the Universe. In a C-H-O system, the elemental abundance can be characterized by the hydrogen abundance (defined by number, and denoted as $X_\ce{H}$) and the carbon versus oxygen ratio (denoted as $X_\ce{C}/X_\ce{O}$). We explore thick atmospheres being hydrogen-rich ($X_\ce{H}\ge0.7$),  hydrogen-intermediate ($0.3< X_\ce{H}<0.7$) and hydrogen-poor ($X_\ce{H}\le0.3$); and we also explore the atmospheres with very different C/O ratios ranging from 0.1 to 10. To be specific, the grid for $X_\ce{H}$ is 0.99, 0.7, 0.5, 0.3, and 0.01; and the grid for $X_\ce{C}/X_\ce{O}$ is 10, 2, 1, 0.5, and 0.1. This parameter exploration is reasonable, considering the masses of the elemental building blocks of the thick atmospheres. Taking GJ~1214~b as an example, the mass of a 1000-bar atmosphere is $10^{-3}$ of the planet's mass. If the atmosphere has $X_\ce{H}=0.5$ and $X_\ce{C}/X_\ce{O}=2$, the mass of hydrogen, carbon, and oxygen in the atmosphere is $7.3\times10^{-5}$, $5.8\times10^{-4}$, $3.9\times10^{-4}$ of the planet's mass. The masses of these building blocks for the thick atmosphere are reasonable.

{\it Incident Stellar Irradiation} The incident stellar irradiation is specified in terms of the irradiation temperature ($T_{\rm irr}$) for the calculation of temperature profiles. The irradiation temperature is defined as $\sigma T_{\rm irr}^4 \equiv \frac{L_{\rm star}}{4\pi a^2}$, where $\sigma$ is the Stefan-Boltzmann constant, $L_{\rm star}$ is the star's luminosity, and $a$ is the planet's semi-major axis. We here explore the molecular composition of thick atmospheres having an irradiation temperature of 770 K, 1000 K, 1200 K, 1400 K, 2000 K, and 2754 K. 770 K is the irradiation temperature of GJ~1214~b, and 2754 K is the irradiation temperature of 55~Cnc~e. For different types of stars that have different luminosities, an irradiation temperature corresponds to different semi-major axes of the simulated planet. For a Sun-like G2V star as the parent star, our grid of irradiation temperature corresponds to semi-major axes of 0.20, 0.12, 0.082, 0.060, 0.030, and 0.016 AU. For an M star like GJ 1214 as the parent star,  our grid of irradiation temperature from 770 K to 1400 K corresponds to semi-major axes of 0.014, 0.0085, 0.0059, and 0.0043 AU. We assume zero planetary albedo and full heat redistribution for the calculation of temperature profiles, so the equilibrium temperature $T_{\rm eq}$ is related to the irradiation temperature as $T_{\rm eq}=(1/4)^{1/4}T_{\rm irr}$. The specific choices on the albedo and the heat redistribution fraction do not affect the generality of our results because they are degenerate with $T_{\rm irr}$ under the grey-atmosphere approximation (Guillot 2010).

{\it Intrinsic Temperature} We assign an intrinsic temperature ($T_{\rm int}$) for the temperature-pressure profile calculation in each scenario to specify the internal heat flux from a planet. The relation between the internal heat flux ($F_{\rm int}$) and the intrinsic temperature is $F_{\rm int} = \sigma T_{\rm int}^4$. The parameterization of  intrinsic temperature has been widely used for simulating temperature-pressure profile of extrasolar gas giants, and appropriate values for the intrinsic temperatures can vary from a few tens K to a few hundreds K depending on the evolution history of the planet  (e.g., Marley et al. 2007; Fortney et al. 2008). For super Earths and mini Neptunes the intrinsic temperatures are unknown and rely on future observations to be determined. We here consider a fiducial intrinsic temperature of 35 K, which is the intrinsic temperature corresponding to the geothermal heat flux. In addition, we explore the effect of an intrinsic temperature being 90 K (similar to the intrinsic temperature for Jupiter) and 180 K for some scenarios.

{\it Stellar Spectral Type} For GJ~1214~b models we use the the latest {\it HST} measurement of the UV spectrum of GJ 1214 (France et al. 2013) and the NextGen simulated visible-wavelength spectrum of an M star having parameters closest to those of GJ 1214 (i.e., effective temperature of 3000 K, surface gravity log($g$)=5.0, and metallicity [M/H] = 0.5; Allard 1997). We also explore of the effect of stellar UV flux by assuming a Sun-like G2V star as the parent star for a GJ~1214~b like exoplanet. For the Sun-like star we use the standard reference solar spectrum (Air Mass Zero) published by the American Society for Testing and Materials\footnote{http://rredc.nrel.gov/solar/spectra/am0/}. For HD~97658~b and 55~Cnc~e models we use a black body spectrum of effective temperature 5200 K with additional Sun-like chromospheric emission in UV wavelengths.

{\it Eddy Diffusivity and Molecular Diffusivity} We explore eddy diffusion coefficients ranging from $10^6$ to $10^9$ cm$^2$ s$^{-1}$, reasonable values for deep atmospheres according to the free-convection and mixing-length theories (Gierasch \& Conrath 1985; Visscher et al. 2010). The eddy diffusion coefficients are assumed to be constant throughout the atmosphere. Such assumption does not consider the possibility of a temperature inversion that may lower the eddy diffusion coefficients by more than 3 orders of magnitude at $\sim0.1$ bar (e.g., Gladstone et al. 1996 for Jupiter). Our models may therefore under-estimate the amounts of potential photochemical products in the upper atmosphere; however this paper is mainly concerned with the transport-driven disequilibrium in the deep atmosphere, for which our assumption regarding the eddy diffusion coefficients is sufficient. In addition to eddy diffusion, our model takes molecular diffusion of \ce{H} and \ce{H2} into account, even though the effect of molecular diffusion for the visible part of the atmosphere ($10^{-4}$-1 Bar) is quite minimal. For an eddy diffusion coefficient ranging from $10^6$ to $10^9$ cm$^2$ s$^{-1}$, the homopause (defined as where molecular diffusion coefficient equates eddy diffusion coefficient) ranges in 0.2-0.0002 Pa for planets as irradiated as GJ~1214~b, and in 2.3-0.0023 Pa for planets as irradiated as 55~Cnc~e. Molecular diffusion is therefore only important for the top layers of the atmospheres on super Earths/mini Neptunes. The effect of molecular diffusion is to enrich H and \ce{H2} and lower the mean molecular mass above the homopause.

%In total, as a general study we have performed 1800 individual photochemistry-thermochemistry simulations to cover wide ranges of controlling parameter for the molecular compositions on super Earths and mini Neptunes. Each simulation invokes iterations between the kinetic-transport model, which itself time-steps to convergence, and the approximated radiative-transfer model. This is to our knowledge by far the most comprehensive computation program executed for the atmosphere of low-mass exoplanets.

\section{Results}

%
%\begin{figure*}[h]
%\begin{center}
% \includegraphics[width=0.8\textwidth]{Compa.eps}
% \caption{C-H-O chemistry of irradiated thick exoplanet atmospheres.
% We have used the stellar and planetary parameters of GJ 1214 system, and assumed an Eddy diffusion coefficient of $10^6$ cm$^{2}$ s$^{-1}$.
% Each panel in the matrix shows chemical composition of simulated atmospheres: the panels in the diagonal show the cases of \ce{H2}, \ce{H2O}, \ce{CH4}, \ce{CO} and \ce{CO2} dominated atmospheres; other panels show the cases of the 50\%-50\% binary combination of the five gases.  Note that not all combination between the five gases is allowed: in particular \ce{H2O} is not compatible with equal amount of \ce{CH4} and \ce{CO}; and \ce{CH4} is not compatible with equal amount of \ce{CO2}.
% }
% \label{Compa}
%  \end{center}
%\end{figure*}

% The main finding is that if a super Earth/mini Neptune has a non-\ce{H2}-dominated thick atmosphere, the major gases in its atmosphere will be \ce{H2O}, \ce{CO2}, \ce{CO}, \ce{CH4}, \ce{C2H4}, \ce{C2H2}, or even \ce{O2}, depending on the hydrogen elemental abundance and the carbon to oxygen elemental abundance ratio. 
We provide a classification scheme of super Earth thick atmospheres based on extensive photochemistry-thermochemistry simulations. The new types of non-\ce{H2}-dominated atmospheres are water-rich atmospheres, oxygen-rich atmospheres, and hydrocarbon-rich atmospheres. We reveal the molecules that could exist in abundance in these types of atmospheres, and outline the means to observationally distinguish atmosphere scenarios via spectral features of hallmark molecules.

\subsection{Chemical Classification of Thick Atmospheres}

\begin{figure*}[h]
\begin{center}
 \includegraphics[width=1.0\textwidth]{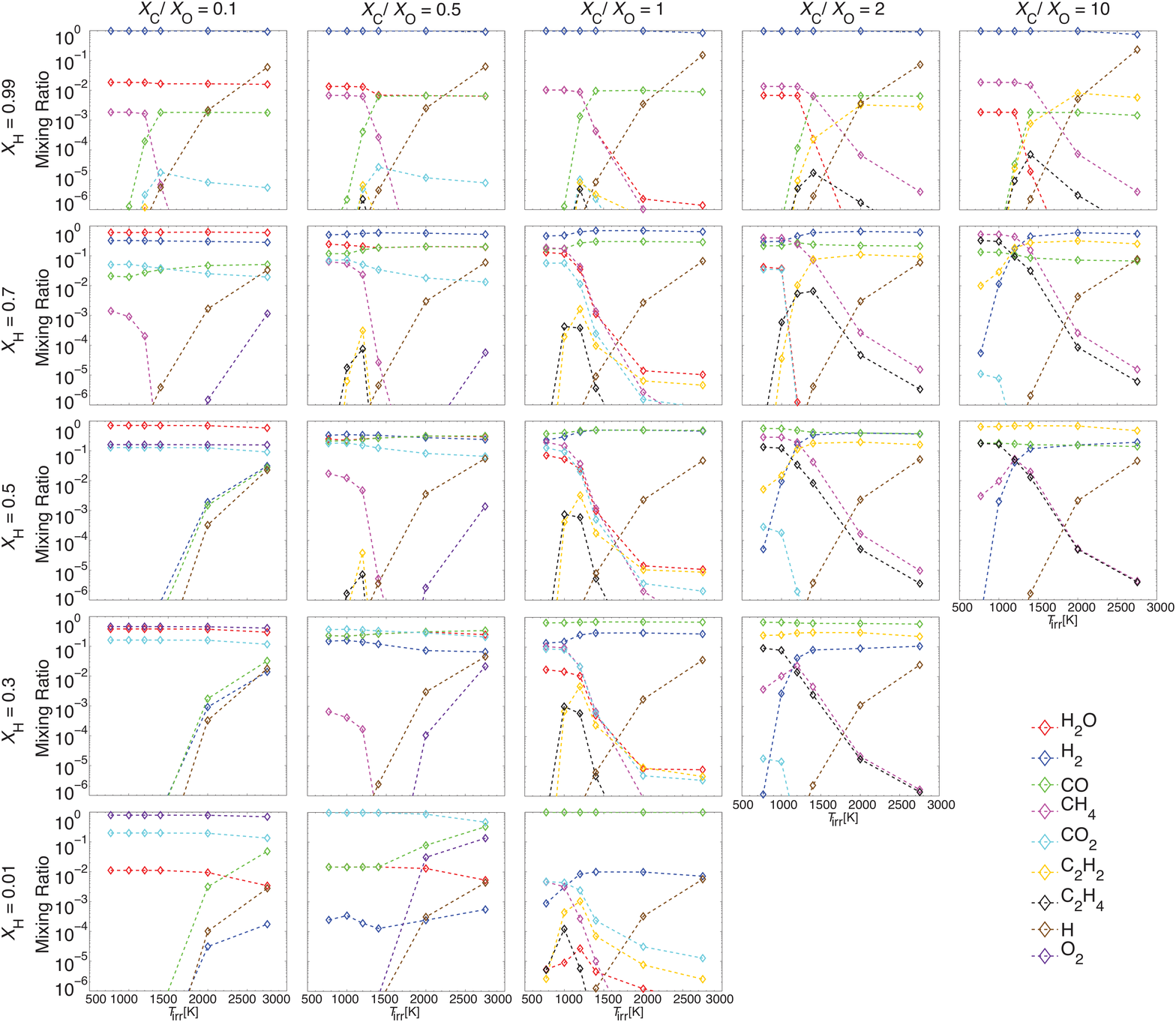}
 \caption{
Mixing ratios of the most important C-H-O species in thick atmospheres as function of the carbon to oxygen elemental abundance ratio and the irradiation temperature ($T_{\rm irr}$). The simulated planet is a 6.5-$M_\oplus$ and 2.7-$R_\oplus$ planet (the same as GJ~1214~b) around an M 4.5 star (for $T_{\rm irr}\le1400$ K) or around a Sun-like star (for $T_{\rm irr}>1400$ K). The eddy diffusion coefficient is assumed to be $10^9$ cm$^2$ s$^{-1}$ and the internal heat flux is assumed to be 35 K. Each diamond marker corresponds to a photochemistry-thermochemistry kinetic-transport simulation with the temperature-pressure profile self-consistently computed. The plotted mixing ratios are the average in the observable pressure levels, i.e., 1-100 mbar. 
%The figure shows that $T_{\rm eq}\sim1000$ K divides the \ce{CH4}-dominated regime and the \ce{CO}-dominated regime. The mixing ratio of \ce{H2O} is much more sensitive to $X_\ce{C}/X_\ce{O}$ in the \ce{CO}-dominated regime than in the \ce{CH4}-dominated regime.
This figure provides a general reference for quickly determining the dominant gases in the thick atmosphere of an exoplanet specified by the temperature, the hydrogen abundance, and the carbon to oxygen abundance ratio.
  }
 \label{General_CHO}
  \end{center}
\end{figure*}

We classify thick atmospheres on super Earths/mini Neptunes broadly into hydrogen-rich atmospheres, water-rich atmospheres, oxygen-rich atmospheres, and hydrocarbon-rich atmospheres, depending on the hydrogen abundance and the carbon to oxygen abundance ratio. The classification is based on the fraction of the gases that have mixing ratios in the order of 0.1 (i.e., the major gases) in the thick atmospheres. If a thick atmosphere is not \ce{H2}-dominated, we find that the dominated C-H-O molecule at the potentially observable levels ($1\sim100$ mbar) can be \ce{H2O}, \ce{CO2}, \ce{CO}, \ce{CH4}, other hydrocarbons (\ce{C2H4} and \ce{C2H2}), or even \ce{O2}. This is the first complete list of major molecular building blocks made of C, H, O elements for thick atmospheres on super Earths and mini Neptunes, and the proposition that unsaturated hydrocarbons including \ce{C2H2} and \ce{C2H4} can be the dominant gases in an exoplanet atmosphere is also the first. Figure \ref{Zoo} summarizes the parameter regimes for each types of atmospheres based on the hydrogen abundance and the carbon to oxygen abundance ratio; Figure \ref{CHO} and Figure \ref{CHO1} show an example of this classification for super Earth/mini Neptune GJ~1214~b and 55~Cnc~e, respectively; and Figure \ref{General_CHO} illustrates the dependency of this classification regime on temperature. These figures show the theoretical range of atmospheric chemical composition; different types of atmosphere may not be equally likely in reality.

Before describing the non-\ce{H2}-dominated atmospheres, we first show that if $X_\ce{H}>0.7$, thick atmospheres of super Earths and mini Neptunes contain abundant \ce{H2} and therefore have the same chemical behaviors as atmospheres of gas giants. Previous thermochemistry and photochemistry results regarding the thick atmospheres on gas giants are valid in these cases, including: \ce{CH4} is the dominant carbon species at equilibrium temperatures lower than $\sim1000$ K; \ce{CO} is the dominant carbon species at equilibrium temperatures higher than $\sim1000$ K (Line et al. 2010; Visscher \& Moses 2011; Madhusudhan 2012); and \ce{H2O} becomes scarce as $X_\ce{C}/X_\ce{O}$ exceeds 1 at the high temperatures that favor \ce{CO} over \ce{CH4}  (e.g., Kuchner \& Seager 2005; Kopparapu et al. 2012; Madhusudhan 2012). We confirm all these behaviors in our simulations for super Earths and mini Neptunes having hydrogen-rich atmospheres (see Figure \ref{General_CHO}). In particular, we observe at the low temperatures where \ce{CH4} is the dominant carbon carrier and \ce{H2O} is the dominant oxygen carrier, the ratio \ce{CH4}/\ce{H2O} is equal to the C/O ratio (Figure \ref{General_CHO}), and in such cases the eddy diffusion coefficients have no effects on the abundances of \ce{CH4} or \ce{H2O} in the observable parts of the atmospheres. At the high temperatures where \ce{CO} is the dominant carbon carrier, the mixing ratio of \ce{H2O} drops dramatically when $X_\ce{C}/X_\ce{O}$ exceeds 1 due to the fact that most oxygen atoms have to be bound with carbon atoms. The exact irradiation temperature for the \ce{CH4}-\ce{CO} transition also depends on the internal heat flux; and the exact mixing ratios of trace species, for example \ce{CO} at the low temperatures and \ce{CH4} at the high temperatures, depend on the strength of eddy mixing and the UV flux of the parent star.

Now we turn to non-\ce{H2}-dominated atmospheres.
%As a super Earth evolves and $X_\ce{H}$ of its atmosphere decreases, new classes of atmospheric compositions emerge. 
If the hydrogen abundance ($X_\ce{H}$) is lower than $0.7\sim0.8$, \ce{H2} is no longer the dominated gas in the atmosphere, and the atmosphere will contain abundant water vapor for low $X_\ce{C}/X_\ce{O}$, and hydrocarbons (i.e., \ce{CH4}, \ce{C2H2}, and \ce{C2H4}) for high $X_\ce{C}/X_\ce{O}$. The water-rich atmospheres occur when $X_\ce{C}/X_\ce{O}\le0.5$ and the hydrocarbon-rich atmospheres occur when $X_\ce{C}/X_\ce{O}\ge2$, for an equilibrium temperature ranging from 500 to 2000 K (Figure \ref{General_CHO}). For the intermediate cases ($0.5<X_\ce{C}/X_\ce{O}<2$), the atmosphere will have abundant \ce{CO} and \ce{H2}, and their fractions are sensitive to $X_\ce{H}$. In addition, \ce{CH4}, \ce{CO2}, and \ce{H2O} can also take a significant fraction and their relative abundances depend on the temperature and the elemental abundance of the atmosphere (Figure \ref{General_CHO}). A general trend in this regime is that \ce{CO} and \ce{C2H2} become more and more abundant and \ce{CH4} and \ce{CO2} become less and less abundant with increasing temperatures (Figure \ref{General_CHO}). If the hydrogen abundance ($X_\ce{H}$) is lower than 0.3 (e.g., for an extremely evolved super Earth on which most atmospheric hydrogen has been lost), significant amounts of \ce{O2} will build up in the atmosphere when $X_\ce{C}/X_\ce{O}<0.5$. We therefore classify super Earth atmospheres according to the hydrogen abundance and the carbon to oxygen abundance ratio, which includes hydrogen-rich atmospheres, water-rich atmospheres, hydrocarbon-rich atmospheres, oxygen-rich atmospheres, and the atmospheres that could have similar amounts of \ce{CO}, \ce{H2}, \ce{CH4}, \ce{CO2}, and \ce{H2O} (Figure \ref{Zoo}).

\begin{figure}[h]
\begin{center}
 \includegraphics[width=0.5\textwidth]{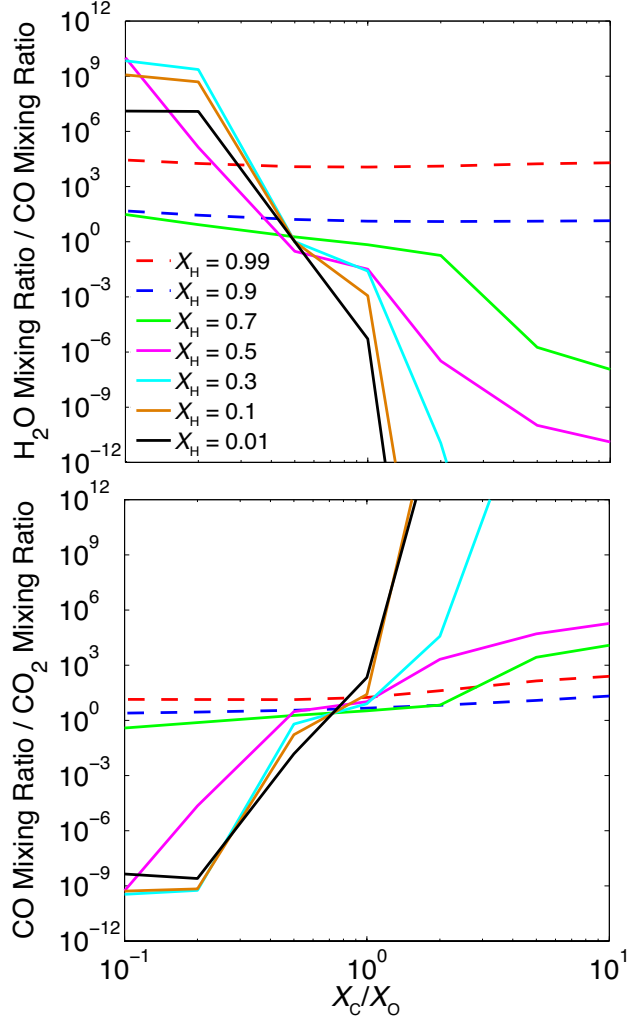}
 \caption{
Key ratios of molecular abundances in thick atmospheres on exoplanets having equilibrium temperatures similar to that of GJ~1214~b as a function of the carbon to oxygen elemental abundance ratio of the atmosphere. Each line with different colors corresponds to a different hydrogen abundance, ranging from hydrogen-rich to hydrogen-poor. The hydrogen rich cases are shown in dashed lines, to be compared with the non-hydrogen-rich cases shown in solid lines.
The $X_\ce{C}/X_\ce{O}$ dependency for non-\ce{H2}-dominated atmospheres is much more sensitive than for \ce{H2}-dominated atmospheres.
  }
 \label{CHO_RATIO}
  \end{center}
\end{figure}

Non-\ce{H2}-dominated atmospheres differ from \ce{H2}-dominated atmospheres mostly in the dependency on the carbon to oxygen abundance ratio. In general, the composition of a non-\ce{H2}-dominated atmosphere is much more sensitive to carbon to oxygen abundance ratio than that of a \ce{H2}-dominated atmosphere (Figure \ref{CHO_RATIO}). This is because when the atmosphere is not hydrogen-rich, the composition is first and foremost constrained by the limited supply of hydrogen. To better illustrate this point, let us consider how the \ce{H2O} versus \ce{CO} abundance ratio depends on $X_\ce{C}/X_\ce{O}$ for an example (see the upper panel of Figure \ref{CHO_RATIO}). For \ce{H2}-dominated atmospheres at the irradiation temperature less than $\sim1200$ K, the main carbon-bearing species is \ce{CH4}, and \ce{CO} is in equilibrium with \ce{H2O}; therefore the \ce{H2O} versus \ce{CO} abundance ratio only has week dependency on $X_\ce{C}/X_\ce{O}$. However, for non-\ce{H2}-dominated atmospheres at similarly low temperatures, \ce{CO} must serve as one of the carbon carriers and then increasing $X_\ce{C}/X_\ce{O}$ will result in more fractions of oxygen to be bound with carbon, and less fractions to be bound with hydrogen to form \ce{H2O}. In all, while the carbon to oxygen abundance ratio affects the amounts of minor gases in \ce{H2}-dominated atmospheres, the carbon to oxygen abundance ratio controls the abundances of major gases in non-\ce{H2}-dominated thick atmospheres.

One might ask how this classification scheme depends on the parameters other than the elemental abundance and the irradiation temperature, e.g., the internal heat flux that  affects the temperature at depth, the eddy diffusivity that determines the efficiency of vertical mixing, and the UV flux that determines the efficiency of photodissociation. Based on our numerical experiments, we find that: (1) the effect of the internal heat flux is to increase or decrease the temperature at the quenching pressure, which to a certain extent leads to the same effect as increasing or decreasing the irradiation level; (2) the effect of changing eddy diffusivity and stellar UV flux in reasonable ranges is not critical for the species that has mixing ratios in the orders of $10^{-1}\sim10^{-2}$, but may be critical for other minor species. The properties presented by Figure \ref{General_CHO} can therefore be considered as general, if the irradiation temperature is regarded as a general proxy for the atmospheric temperature that could also contain contribution from internal heating. Because the effects of the internal heat flux, the eddy diffusivity, and the stellar spectrum on minor species depend on specific atmospheric scenarios, we defer the discussion of that aspect to the subsequent sections dedicated to specific types of atmospheres.

Interestingly, for certain endmember scenarios of non-\ce{H2}-dominated thick atmospheres, we find that the elemental abundance uniquely determines the amounts of major components in the atmosphere for a wide range of temperatures, and we derive analytical formulae of the molecular compositions for those scenarios. Table \ref{Para} tabulates an exhaustive list of parameter regimes of the elemental abundance, and the derived formulae for mixing ratios of major components in these regimes. The fundamental principle that we use here is simply that all H, C, and O atoms in the atmosphere have to be bound to form a limited set of thermochemically stable molecules (e.g., \ce{H2}, \ce{H2O}, \ce{CO2}, \ce{CO}, \ce{CH4}, \ce{C2H4}, \ce{C2H2}, \ce{O2}, etc.), to the extent that molecular forms of C, H, and O elements are thermochemically preferred over atomic forms (which typically corresponds to an equilibrium temperature of less than $\sim2500$ K) and that water vapor does not condense in the atmosphere (which typically corresponds to an equilibrium temperature of greater than $\sim300$ K). In the cases of $X_\ce{C}\ll1$ and the cases of $X_\ce{H}\ll1$, the abundances of major gases in the atmosphere are uniquely determined by the elemental abundance (Table \ref{Para}). 

Finally, we outline the key molecules as remote-sensing probes for the atmospheric scenarios of super Earths and mini Neptunes as a cookbook for observers. Note that molecules such as \ce{H2O} and \ce{CO2} can typically produce considerable features in an exoplanet spectrum even for very low abundances. For the purpose of the cookbook, we apply a mixing ratio threshold of $10^{-6}$ for considering \ce{H2O}, \ce{CO2}, and \ce{CH4} to be abundant, and $10^{-4}$ for \ce{CO} and hydrocarbons. The cookbook is mainly summarized from Figure \ref{General_CHO}, and the most significant effects of internal heating and photolysis are also included.
\begin{itemize}
\item \ce{H2O} is generally abundant in any thick atmosphere that has $X_\ce{C}/X_\ce{O}\le1$ and an irradiation temperature ranging from 700 to 2700 K. 
%\ce{H2O} is also abundant for $X_\ce{C}/X_\ce{O}\sim1$ only when $T_{\rm irr}\le1200$ K, and \ce{H2O} abundance quickly drops off at higher temperatures due to loss of oxygen to \ce{CO} from the \ce{CH4 <=> CO} conversion. 
\ce{H2O} is not expected to have an observable abundance (i.e., mixing ratio greater than $10^{-6}$) in any thick atmospheres that has $X_\ce{C}/X_\ce{O}$ significantly greater than unity, except for those dominated by \ce{H2} and having $T_{\rm irr}\le1200$ K.
\item The parameter space for which \ce{CO2} is abundant (i.e., mixing ratio greater than $10^{-6}$) is similar to the parameter space for \ce{H2O}; in other worlds \ce{CO2} and \ce{H2O} should be generally expected to coexist. The exceptional parameter regime is \ce{H2}-dominated atmospheres with $X_\ce{C}/X_\ce{O}\le0.5$ and $T_{\rm irr}\le1200$ K; in this regime the atmospheres have abundant \ce{H2O} but not \ce{CO2}.
\item \ce{CO} is expected to be abundant (i.e., mixing ratio greater than $10^{-4}$) in any thick atmosphere that has $X_\ce{H}\le0.7$, $X_\ce{C}/X_\ce{O}\ge0.5$, and an irradiation temperature ranging from 700 to 2700 K. Otherwise \ce{CO} is not expected to be abundant in the atmosphere, unless the irradiation temperature is greater than 1200 K.
\item \ce{CH4} is not expected to have an observable abundance (i.e., mixing ratio greater than $10^{-6}$) in any atmosphere that has $T_{\rm irr}>1400$ K (or, significant internal heating with $T_{\rm int}>200$ K) and $X_\ce{C}/X_\ce{O}\le1$, and \ce{CH4} is generally abundant at cooler temperatures. The only scenario in which $T_{\rm irr}\le1400$ K and yet methane is not abundant is the \ce{H2O}/\ce{CO2}-dominated atmosphere.
\item Hydrocarbons (i.e., \ce{C2H2} and \ce{C2H4}) are expected to be abundant (i.e., mixing ratio greater than $10^{-4}$) in any atmospheres that has $X_\ce{H}\le0.7$, $X_\ce{C}/X_\ce{O}\ge2$, and an irradiation temperature ranging from 700 to 2700 K. Abundant hydrocarbons can also be expected in the atmospheres that have abundant \ce{CH4} and efficient photolysis.
\end{itemize}

% When an atmosphere is hydrogen-rich, supply of hydrogen atom is virtually unlimited, which allows C and O to be bound with as much H as needed to achieve the globally minimized Gibbs free energy at depth (i.e., thermochemical equilibrium). In contrast, when the atmosphere is not hydrogen-rich, the supply of hydrogen is limited, and the atmospheric composition is first and foremost subject to this constraint. 

In the following two subsections, we present the results for the two new types of thick atmospheres on super Earths and mini Neptunes: water-rich atmospheres and hydrocarbon-rich atmospheres. Oxygen-rich atmospheres are described together with water-rich atmospheres as they both correspond to low $X_\ce{C}/X_\ce{O}$ scenarios. We focus on the ranges of possible molecular compositions in these atmospheres, because it is the molecular compositions that control the observational properties of these atmospheres.

\subsection{Ranges of Compositions of Water-Rich Atmospheres and Oxygen-Rich Atmospheres}

%We here present the results regarding water-rich and oxygen-rich thick atmospheres on super Earths and mini Neptunes. 
Water-rich atmospheres are in particular interesting because water is a substance fundamental for life. We have already shown that water-rich atmospheres emerge as an atmosphere with low $X_\ce{C}/X_\ce{O}$ loses its most of its free hydrogen (Figure \ref{Zoo}). For low $X_\ce{C}/X_\ce{O}$, the atmosphere is a mixture of \ce{H2} and \ce{H2O} when $X_\ce{H}\geq 2X_\ce{O}$ and \ce{H2O} and \ce{O2} when $X_\ce{H}< 2X_\ce{O}$ (Table \ref{Para}). In this section, we study the possible ranges of mixing ratios for both the main components and the minor components in water-rich atmospheres. The main components are important because they are directly controlled by the elemental abundances and they serve as the background atmosphere; and the minor components are also important because they may lead to significant, if not dominant, spectral features and they provide means to characterize vertical mixing and internal heating of a super Earth exoplanet. 

% The major components in a water-rich atmosphere on a super Earth are \ce{H2}, \ce{H2O}, and \ce{O2}. The water-rich atmosphere corresponds to the parameter regime of $X_\ce{H}\sim1$, $X_\ce{O}\sim1$, $X_\ce{C}\ll1$. As tabulated in Table \ref{Para}, when $X_\ce{H}\geq 2X_\ce{O}$, the atmosphere is a mixture of \ce{H2} and \ce{H2O}; when $X_\ce{H}< 2X_\ce{O}$, the atmosphere is a mixture of \ce{H2O} and \ce{O2}. This is true regardless the atmospheric temperature, as long as the three molecules (\ce{H2}, \ce{H2O}, and \ce{O2}) rather than atomic hydrogen or oxygen are thermochemically preferred. As a super Earth evolves and loses most of free hydrogen in its atmosphere, and its atmosphere may become water-dominated. Subsequent loss of hydrogen from water would lead to accumulation of free oxygen in the atmosphere.

\begin{figure}[h]
\begin{center}
 \includegraphics[width=1.0\textwidth]{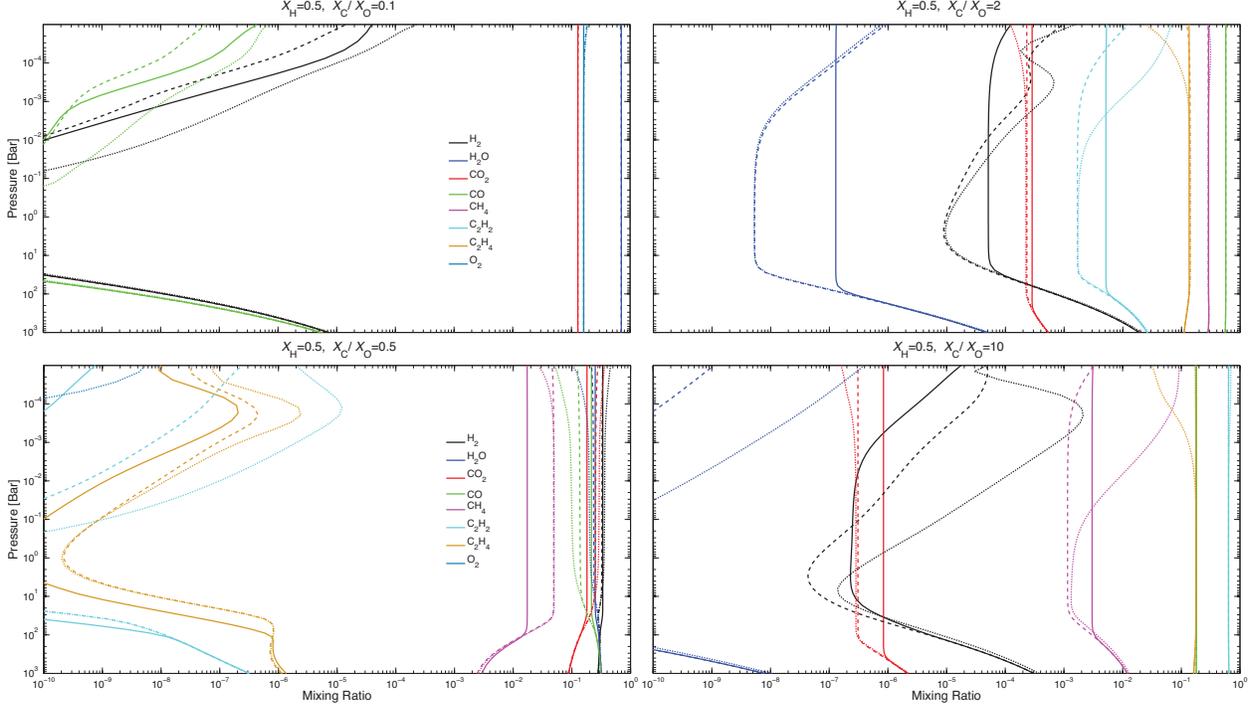}
 \caption{
 The effect of eddy diffusivity and stellar spectrum on the molecular composition of water-rich and hydrocarbon-rich atmospheres on GJ~1214~b like super Earths/mini Neptunes. The simulated super Earth atmosphere has an adopted hydrogen abundance of 0.5 and a carbon to oxygen abundance ratio ranging from 0.1 to 10. The planet is a GJ~1214~b sized planet at the 0.014-AU orbit around an M4.5 star. 
 %With an internal temperature of 20 K, the planet has a temperature of 470 K at the top of atmosphere, a temperature of $\sim810$ K (top) and $\sim830$ K (bottom) at the pressure level of 1 bar, and a temperature of $\sim1300$ K at the pressure level of 100 bar.  
 Results are shown for a vertical eddy diffusion coefficient of $10^9$ cm$^2$ s$^{-1}$ (solid lines) and $10^6$ cm$^2$ s$^{-1}$ (dashed lines). Results are also shown for a vertical eddy diffusion coefficient of $10^6$ cm$^2$ s$^{-1}$ and a Sun-like star as the parent star (dotted lines).
 With $X_\ce{C}/X_\ce{O}=0.1$, water vapor is the most abundant gas in the observable atmosphere with a mixing ratio of 0.71, and \ce{O2} and \ce{CO2} are the next abundant gases with mixing ratios of 0.16 and 0.13; whereas with a solar $X_\ce{C}/X_\ce{O}$, \ce{H2}, \ce{CO}, \ce{CO2}, and \ce{H2O} all have mixing ratio in the order of 0.1 at the pressure level of $1\sim100$ mbar, and the exact composition is sensitive to the eddy diffusion coefficient and the stellar input spectrum. 
 For $X_\ce{C}/X_\ce{O}=2$, \ce{CO}, \ce{CH4}, and \ce{C2H4} have mixing ratios in the order of 0.1 at the pressure level of $1\sim100$ mbar; and for $X_\ce{C}/X_\ce{O}=10$, \ce{C2H2} and \ce{C2H4} have mixing ratios in the order of 0.1.
  }
 \label{CHEM_KZZ}
  \end{center}
\end{figure}

We first consider the main components of water-rich atmospheres. Among the water-rich atmospheres, water-dominated atmospheres are defined as the atmospheres in which water is the most abundant gas. Our atmosphere models show that water-dominated atmospheres only occur when $X_\ce{H}\sim2X_\ce{O}$ and $X_\ce{C}/X_\ce{O}\ll1$, a fairly small part of the parameter space of elemental abundances. Mixture of carbon, even at the solar $X_\ce{C}/X_\ce{O}$, would lead to the removal of most atmospheric water. As shown in Figure \ref{CHEM_KZZ}, when $X_\ce{H}=0.5$, $X_\ce{C}/X_\ce{O}=0.1$, the atmosphere is dominated by water, with a trace amount of carbon in the form of \ce{CO2}.  The water vapor abundance decreases dramatically as the atmosphere has more carbon (Figure \ref{CHO_RATIO}). 
%Especially for non-hydrogen-rich atmospheres the water vapor become extinctive if the C/O ratio is greater than some cut-off value ranging from 1 to 3 (see Figure \ref{CHO}). On the other side, the water vapor mixing ratio is always greater than $10^{-3}$ for a C/O ratio smaller than 0.5 (Figure \ref{CHO}).
At the solar $X_\ce{C}/X_\ce{O}$, the atmosphere would be a mixture of \ce{H2}, \ce{CO}, \ce{CO2}, \ce{CH4}, and \ce{H2O}, in which the water vapor mixing ratio is about 10\% (Figure \ref{CHEM_KZZ}). Such an atmosphere is not water-dominated, but such an atmosphere shall still be considered water-rich, because a water vapor amount of even less than 10\% would be very significant in the spectrum. A conceptual way to understand this sensitivity of water abundance on the carbon content is that for a wide temperature range, oxygen atoms tend to be bound with carbon atoms whenever available. In summary, water-rich atmospheres exist when $X_\ce{C}/X_\ce{O}\leq0.5$, but water-dominated atmospheres exist only when $X_\ce{C}/X_\ce{O}\ll0.5$.

\begin{figure}[h]
\begin{center}
 \includegraphics[width=0.5\textwidth]{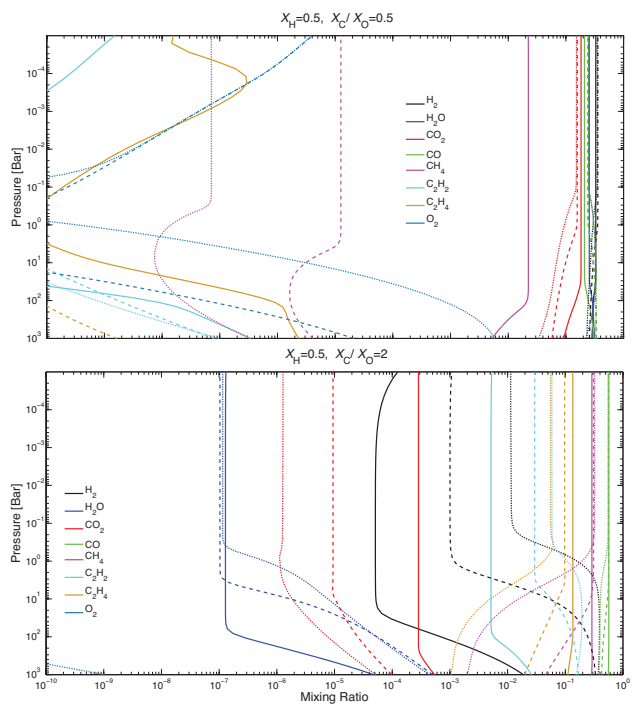}
 \caption{
 The effect of internal heat flux on the molecular composition of water-rich and hydrocarbon-rich atmospheres on GJ~1214~b like super Earths/mini Neptunes. The simulated atmosphere has an adopted hydrogen abundance of 0.5 and a carbon to oxygen abundance ratio of 0.5 and 2. The planet is a GJ~1214~b sized planet at the 0.014-AU orbit around an M4.5 star. 
 %With an internal temperature of 20 K, the planet has a temperature of 470 K at the top of atmosphere, a temperature of $\sim810$ K (top) and $\sim830$ K (bottom) at the pressure level of 1 bar, and a temperature of $\sim1300$ K at the pressure level of 100 bar.  
 Results are shown for a vertical eddy diffusion coefficient of $10^9$ cm$^2$ s$^{-1}$, and an internal heat flux of 35 K (solid lines), 90 K (dashed lines), and 180 K (dotted lines).
 The internal heat flux has little effect on the mixing ratios of the species whose mixing ratios are greater than 0.1 (i.e., major species) in the observable part of the atmosphere, but may be very important for other species in equilibrium with those major species. A general trend is that for a greater internal heat flux, the observable part of atmosphere would have less \ce{CO2} and \ce{CH4}, and more \ce{H2} and \ce{C2H2}.
  }
 \label{CHEM_INT}
  \end{center}
\end{figure}

%While the mixing ratios of main components in the water-dominated atmospheres are uniquely determined by the elemental abundance, 
The composition of the water-rich but not water-dominated atmospheres can be affected by the temperature, the efficiency of vertical mixing, and the stellar input spectrum (see Figure \ref{General_CHO}, \ref{CHEM_KZZ}, and \ref{CHEM_INT}). The mixing ratio of \ce{CH4} is highly sensitive to the irradiation level and the internal heat flux. With an irradiation temperature  of 770 - 1200 K and a small intrinsic temperature of 35 K, the mixing ratio of \ce{CH4} can be as high as $10^{-2}$ in a water-rich but not water-dominated atmosphere. Increasing the irradiation temperature to above 1200 K or the intrinsic temperature from 35 K to 90 K would decrease the \ce{CH4} mixing ratio by several orders of magnitude (Figure \ref{General_CHO} and \ref{CHEM_INT}). Therefore \ce{CH4} can be an effective probe for the temperature in the deep atmosphere. Other major constituents that have mixing ratios greater than 0.1 are not significantly affected by the temperature. The mixing ratios of \ce{CH4} and \ce{CO} can increase by a factor of a few and the mixing ratios of \ce{CO2} and \ce{H2O} can decrease by a few tens percent when the eddy diffusion coefficient decreases from $10^9$ to $10^6$ cm$^2$ s$^{-1}$ (Figure \ref{CHEM_KZZ}). Interestingly, using the solar spectrum that gives considerably more near UV fluxes than does an M dwarf spectrum, a significant fraction of \ce{H2O} and \ce{CO} can be converted into \ce{CO2} and \ce{H2}, while the atmosphere still remains a roughly equal mixture of \ce{H2}, \ce{CO}, \ce{CO2}, and \ce{H2O} (Figure \ref{CHEM_KZZ}).

% Interestingly, the solar C/O ratio is the transition between water-rich and water-poor, and with the solar C/O ratio, all major forms of carbon including \ce{CH4}, \ce{CO}, \ce{CO2}, and water vapor can co-exist in the atmospheres with significant abundance (mixing ratio greater than 0.01; see Figure \ref{CHEM_KZZ} for an example) In this case, the exact composition depends on the effects of eddy mixing. This result shows that at the temperature of GJ~1214~b like super Earths ($\sim1000$ K at depth), oxygen tends to bound with carbon rather than hydrogen. The water vapor abundance in super Earth thick atmospheres critically depends on the atmospheric C/O ratio. Only when the atmosphere has much more oxygen than carbon can water be formed in abundance. 

\begin{figure}[h]
\begin{center}
 \includegraphics[width=0.5\textwidth]{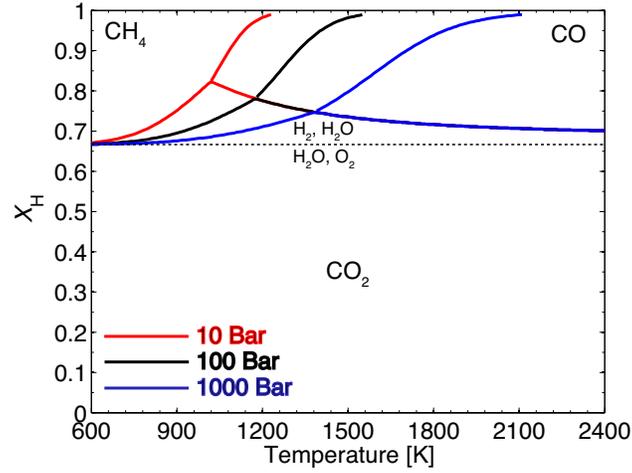}
 \caption{
 Carbon speciation in water-dominated atmospheres on super Earths. The main components of the atmosphere are either \ce{H2} and \ce{H2O} ($X_\ce{H}>2/3$), or \ce{H2O} and \ce{O2} ($X_\ce{H}<2/3$), and their mixing ratios depend on the hydrogen abundance $X_\ce{H}$ as tabulated in Table \ref{Para}. The figure shows the parameter regimes in which \ce{CH4}, \ce{CO}, and \ce{CO2} is the main carbon carrier in the atmosphere. The boundaries between the regimes are computed using equations (\ref{s1}-\ref{send}), for a variety of pressures shown by different colors. As long as the atmosphere is depleted in molecular hydrogen, the main form of carbon is \ce{CO2} regardless of temperature and pressure.
  }
 \label{WATER_T}
  \end{center}
\end{figure}

We now turn to consider the minor components in the water-dominated atmospheres. The minor components of interest are, of course, the three stable forms of carbon, \ce{CH4}, \ce{CO}, and \ce{CO2}. We here outline an analytical treatment for the carbon speciation in the water-dominated atmospheres, which is also applicable to study the thermochemical speciation of other elements in a given mixture of gases. When $X_\ce{H}\sim1$, $X_\ce{O}\sim1$, $X_\ce{C}\ll1$, we may consider carbon as perturbation on a background atmosphere made of either \ce{H2}, \ce{H2O}, or \ce{H2O}, \ce{O2} (Table \ref{Para}).  For carbon in a \ce{H2}-\ce{H2O} system, we have the following balanced reactions:
\reaction{H2O + CH4 <=> 3H2 + CO , \label{H2_CH4CO}}
\reaction{H2O + CO <=> H2 + CO2 . \label{H2_COCO2}}
Let $K_1$ be the equilibrium constant of reaction (\ref{H2_CH4CO}), and $K_2$ be the equilibrium constant of reaction (\ref{H2_COCO2}). Note that $K_1$ and $K_2$ only depend on the temperature. The {\it Law of Mass Action} reads
\begin{eqnarray}
&&\frac{X_\ce{CO}}{X_\ce{CH4}} = \frac{K_1}{P^2}\frac{X_\ce{H2O}}{X_\ce{H2}^3}, \label{s1}\\
&& \frac{X_\ce{CO2}}{X_\ce{CO}} = K_2\frac{X_\ce{H2O}}{X_\ce{H2}},
\end{eqnarray}
in which $X$ denotes the mixing ratio of a molecule, and $P$ is the atmospheric pressure in the unit of bar. Similarly for a \ce{H2O}-\ce{O2} system,
\reaction{3O2 + 2CH4 <=> 4H2O + 2CO , \label{O2_CH4CO}}
\reaction{O2 + 2CO <=> 2CO2 . \label{O2_COCO2}}
We have also
\begin{eqnarray}
&& \frac{X_\ce{CO}}{X_\ce{CH4}} = \sqrt{\frac{K_3}{P}\frac{X_\ce{O2}^3}{X_\ce{H2O}^4}},\\
&& \frac{X_\ce{CO2}}{X_\ce{CO}} = \sqrt{ K_4PX_\ce{O2} }, \label{send}
\end{eqnarray}
in which $K_3$ is the equilibrium constant of reaction (\ref{O2_CH4CO}), and $K_4$ is the equilibrium constant of reaction (\ref{O2_COCO2}). The relative abundance of \ce{CH4}, \ce{CO}, and \ce{CO2} in thermochemical equilibrium with a predominantly hydrogen and oxygen atmosphere can be calculated with equations (\ref{s1}-\ref{send}), for a variety of temperatures and pressures. Figure \ref{WATER_T} summarizes the result.

We find that when the atmosphere is mainly a \ce{H2}-\ce{H2O} mixture ($X_\ce{H}>2X_\ce{O}$), the main form of carbon is either \ce{CH4} at low temperatures or \ce{CO} at high temperatures (similar to gas giant atmospheres); when the atmosphere is mainly a \ce{H2O}-\ce{O2} mixture ($X_\ce{H}<2X_\ce{O}$), the main form of carbon is always \ce{CO2}, regardless of temperature (Figure \ref{WATER_T}; also see Figure \ref{CHEM_KZZ} for an example). The transition from \ce{CH4} to \ce{CO} in a \ce{H2}-\ce{H2O} atmosphere depends on the temperature at the quenching pressure (i.e., the pressure at which the eddy mixing timescale is equal to the thermochemical equilibrium timescale) and the hydrogen abundance (see Figure \ref{WATER_T}). Therefore the abundances of \ce{CH4} and \ce{CO} can probe the efficiency of eddy mixing and the flux of internal heating. A higher quenching pressure (in other words more efficient eddy mixing) would lead to more \ce{CH4} and less \ce{CO} in the observable atmosphere. When the atmosphere is depleted in molecular hydrogen, however, the dominant form of carbon is always \ce{CO2} to the extent that carbon is a minor constituent in the atmosphere ($X_\ce{C}/X_\ce{O}\ll0.5$). In this case, the carbon speciation is insensitive to the effect of eddy mixing or the temperature structure of the atmosphere, and therefore cannot serve as the probe to these physical quantities. In particular, \ce{CH4} is thermochemically prohibited to exist in abundance in a \ce{H2}-depleted water-dominated atmosphere; therefore, nonexistence of \ce{CH4} can be an indicator for water dominance in the atmosphere.

Finally, we comment on the rise of free oxygen in the water-rich atmospheres. Previous discussions have shown that when $X_\ce{H}<2X_\ce{O}$, one would expect the atmosphere to be a mixture of \ce{H2O}, \ce{O2}, and \ce{CO2} (see Figure \ref{CHO}). The free oxygen in the atmosphere is basically the left-over oxygen after forming \ce{H2O} and \ce{CO2}; in other words, free oxygen is expected when $X_\ce{O}>0.5X_\ce{H}+2X_\ce{C}$. The rise of free oxygen we present here is a result of the loss of hydrogen, rather than the photon-driven processes that only affects the upper atmosphere at $\sim10^{-5}$ bar (see the upper panel of Figure \ref{CHEM_KZZ} for an example). The caveat here is that we do not consider any material exchange between the atmosphere and the surface (if there is a surface). The oxygen build-up from hydrogen loss could be prevented by active volcanic release of reduced gases or oxidative weathering of the surface, a process that has been proposed to have operated on Venus (e.g., Kasting 1997). A side point is that the \ce{OH} radical would be very abundant in such \ce{H2O}-\ce{O2}-\ce{CO2} atmospheres, with mixing ratios varying between $10^{-8}$ and $10^{-4}$. As \ce{OH} radicals remove most other gases in the atmosphere rapidly, such a high concentration of \ce{OH} necessarily implies that the any gas that reacts with \ce{OH} without reforming pathways is not expected to be able to accumulate in the atmosphere. The \ce{H2O}-\ce{O2}-\ce{CO2} atmospheres on super Earths are likely to be highly oxidized, with all other elements in their most oxidized form.

%For another example, when H is extremely rare in the atmosphere (H$\leq$0.1) and the atmospheric C/O is smaller than that of \ce{CO2} (0.5), oxygen atoms in excess to forming \ce{CO2} can only form molecular oxygen,  and thus oxygen-rich atmospheres. 

%Molecular oxygen becomes one of the main component in the atmosphere when $X_\ce{H}<0.5$ and $X_\ce{C}/X_\ce{O}<0.5$. The oxygen atom is favorably bound with carbon, and a C/O ratio smaller than 0.5 ensures that there are significant amounts of oxygen atom left over after forming \ce{CO2}, the most oxidized form of carbon. 

\subsection{Ranges of Compositions of Hydrocarbon-Rich Atmospheres}

\begin{figure*}[h]
\begin{center}
 \includegraphics[width=1.0\textwidth]{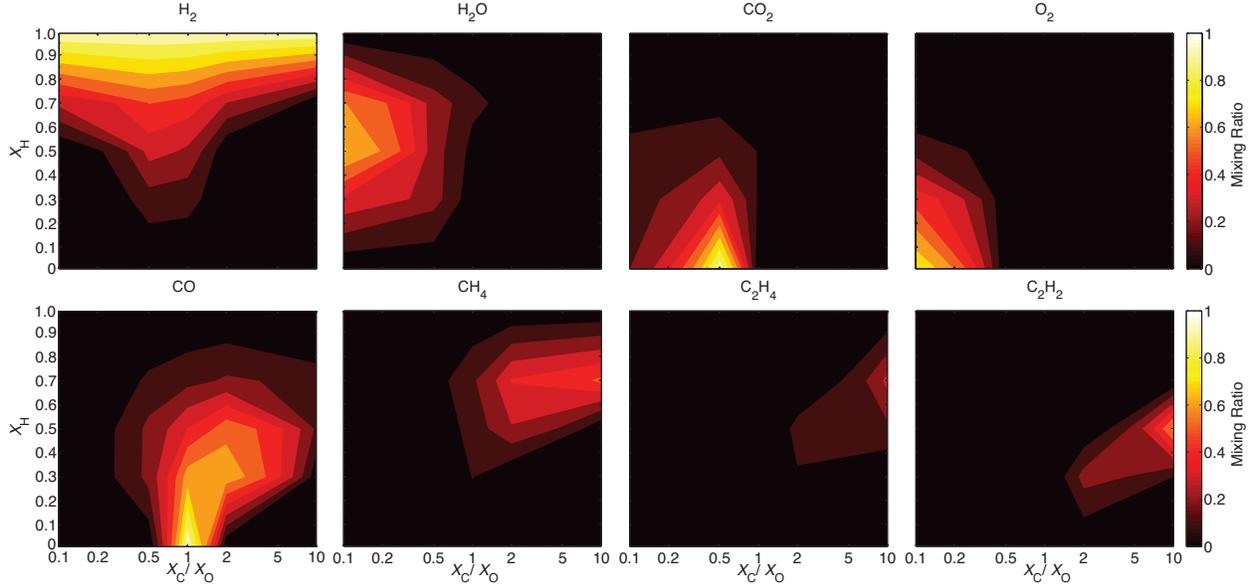}
 \caption{
 Mixing ratios of common molecules in thick atmospheres on a GJ~1214~b like exoplanet. The simulated planet is a 6.5-$M_\oplus$ and 2.7-$R_\oplus$ planet at the 0.014-AU orbit of an M 4.5 star, corresponding to GJ~1214~b. The mixing ratios of \ce{H2}, \ce{H2O}, \ce{CO}, \ce{CO2}, \ce{O2}, \ce{CH4}, \ce{C2H4}, and \ce{C2H2} are shown as a function of the H abundance and the C/O ratio in the atmospheres. % The molecular compositions and the temperature structures of the atmospheres are consistently computed using the photochemistry-thermochemistry model and the grey-atmosphere radiative-convective model. 
 For an irradiation temperature of $T_{\rm irr}=770$ K, the atmospheres have a temperature of $460\sim470$ K at the top, $840\sim1020$ K at 1 bar, and $1500\sim2000$ K at 1000 bar self-consistently computed with the composition assuming $T_{\rm int} = 35$ K. For photochemical calculations, we have used the latest HST measurement of UV flux of GJ 1214 (France et al. 2013), and explored the eddy diffusion coefficients ranging from $10^6$ to $10^9$ cm$^2$ s$^{-1}$. 
 The mixing ratio shown in the figure is the vertically averaged mixing ratio for pressure levels from 1 to 100 mbar, to which transmission and emission spectroscopy is sensitive. Water should not exist in the atmosphere with substantial amounts if $X_\ce{C}/X_\ce{O}>1$ for a wide range of hydrogen abundance; hydrocarbons (\ce{CH4} and \ce{C2H_x}) have high abundances in carbon-rich atmospheres; and molecular oxygen appears abundantly in the atmosphere only for hydrogen-poor and carbon-poor cases. 
  }
 \label{CHO}
  \end{center}
\end{figure*}

\begin{figure*}[h]
\begin{center}
 \includegraphics[width=1.0\textwidth]{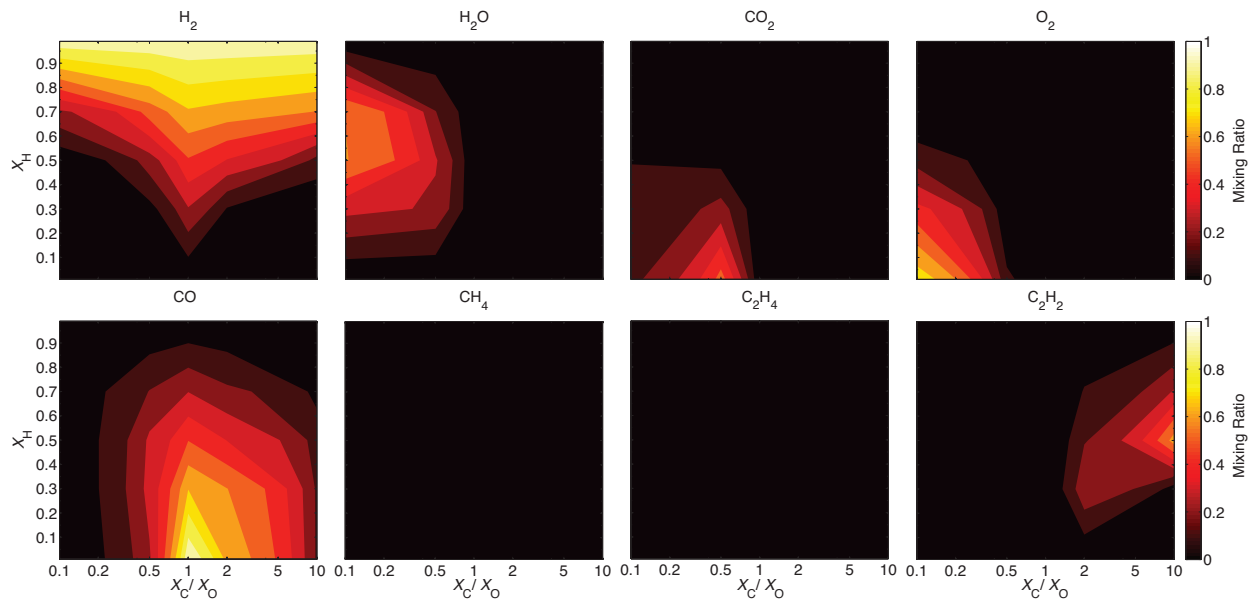}
 \caption{
 Mixing ratios of common molecules in thick atmospheres on a 55~Cnc~e like exoplanet. The same format as Figure \ref{CHO}, but for a 8.6-$M_\oplus$ and 2.0-$R_\oplus$ planet at the 0.0156-AU orbit of a K0V star having an effective temperature of 5200 K, corresponding to 55~Cnc~e.
%The atmospheres have a temperature of 2000-2200 K at the top, and 2500-3000 K at 100 bar self-consistently computed with the composition assuming the interior heat flux to be $T_{\rm int} = 35$ K. 
 Compared with the GJ~1214~b scenarios shown in Figure \ref{CHO}, \ce{CH4} and \ce{C2H4} are not expected to exist in abundance in the atmosphere of 55~Cnc~e; \ce{CO} is the dominant gas in the atmosphere for a much wider parameter space; and \ce{C2H2} becomes the thermochemically preferred hydrocarbons in the atmosphere. 
  }
 \label{CHO1}
  \end{center}
\end{figure*}

Another new type of super Earth atmosphere we find is hydrocarbon-rich atmosphere. The hydrocarbon-rich atmospheres are the conjugate situation of the water-rich atmospheres; the difference here is that while the only stable ``hydro-oxygen" is \ce{H2O}, there are many stable hydrocarbons. Based on our simulations, the potentially observable parts of a hydrocarbon-rich atmosphere may be composed of \ce{H2}, \ce{H}, \ce{CH4}, \ce{C2H2}, \ce{C2H4}, and other higher-order hydrocarbons (Figures \ref{CHO} and \ref{CHO1}). As the number of possible molecules in hydrocarbon-rich atmospheres exceeds the number of major elements in the atmosphere (2), the main components are not uniquely set by the elemental abundance, and depend on other factors including the efficiency of vertical mixing and the temperatures in the atmosphere.

% For example, when H is limited, not all carbon atoms can be bound with 4 hydrogen atoms to form methane; such limitation of hydrogen would lead to hydrocarbon-rich atmospheres, in which one carbon atom is paired with much less hydrogen atoms to form unsaturated hydrocarbons. 

% At the temperature of GJ~1214~b, we find that carbon is in the most reduced form, \ce{CH4}, when the atmosphere has free hydrogen (i.e., $X_\ce{H}>4X_\ce{C}$). If $X_\ce{H}<4X_\ce{C}$, the abundance of hydrogen in the atmosphere is not enough to saturate carbon, and \ce{C2H4}, \ce{C2H2}, and even \ce{C2H} emerge as the main carbon carrier for decreasing hydrogen abundances (see Figure \ref{CHO} and Figure \ref{HYDROCARBON_T}). We here find that, for an exoplanet having carbon-rich atmosphere, hydrocarbons will be formed as the planet loses its hydrogen in the atmosphere. At the temperature of 55~Cnc~e, we find that \ce{CH4} and \ce{C2H4} are not expected to exist in abundance in the atmosphere in any cases, instead \ce{C2H2} and \ce{C2H} would be the main carbon carrier (see Figure \ref{CHO1} and Figure \ref{HYDROCARBON_T}).

\begin{figure}[h]
\begin{center}
 \includegraphics[width=0.5\textwidth]{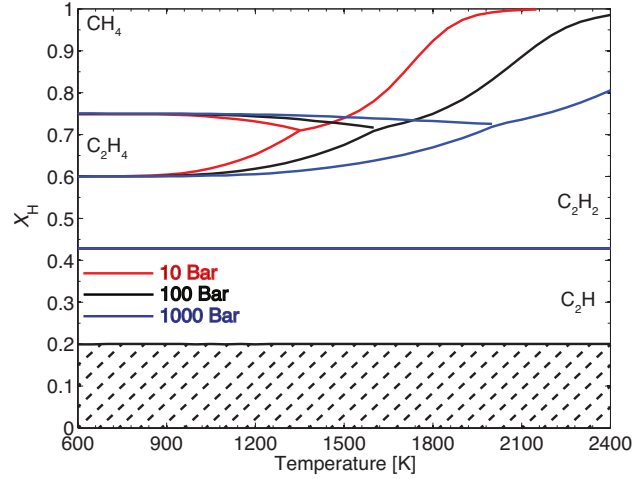}
 \caption{
 Carbon speciation in hydrocarbon-dominated atmospheres on super Earths. The atmosphere has hydrogen and carbon, and very little oxygen, i.e., $X_\ce{H}+X_\ce{C} = 1$ and $X_\ce{O}\ll 1$. We compute the molecular composition in thermochemical equilibrium, and show on the figure the parameter regimes in which \ce{CH4}, \ce{C2H4}, \ce{C2H2}, and \ce{C2H} is the main carbon carrier in the atmosphere. The boundaries between the regimes are for a variety of pressures shown by different colors. The shaded parameter regime ($X_\ce{H}<0.2$) is not physically plausible because it corresponds to the formation of elemental carbon, which would condense and precipitate.
 The scarcity of hydrogen is the main driver for the formation of unsaturated hydrocarbons.
  }
 \label{HYDROCARBON_T}
  \end{center}
\end{figure}

For a super Earth/mini Neptune having a carbon-rich atmosphere, unsaturated hydrocarbons can become the dominant carbon-bearing gases in the atmosphere if the atmosphere loses most of its free hydrogen. When the atmosphere has free hydrogen (i.e., $X_\ce{H}>4X_\ce{C}$), carbon is in the most reduced form, \ce{CH4}, if the planet receives similar degree of irradiation as GJ~1214~b (see Figure \ref{CHO} and Figure \ref{HYDROCARBON_T}). If $X_\ce{C}<X_\ce{H}<4X_\ce{C}$, the abundance of hydrogen in the atmosphere is not enough to saturate carbon, and \ce{C2H4} and \ce{C2H2} emerge as the main carbon carrier for decreasing hydrogen abundances (see Figure \ref{CHO} and Figure \ref{HYDROCARBON_T}). If the planet receives similar degree of irradiation as 55~Cnc~e, \ce{CH4} and \ce{C2H4} are not expected to exist in abundance in the atmosphere in any cases as they are thermochemically unstable at high temperatures, instead \ce{C2H2} would be the main carbon carrier (see Figure \ref{CHO1} and Figure \ref{HYDROCARBON_T}). For highly evolved carbon-rich atmospheres with $X_\ce{H}<X_\ce{C}$, high-order hydrocarbons with more than 2 carbon atoms should form in the atmosphere. Our chemistry model stops at \ce{C2H_x} and produces abundant \ce{C2H} in the atmosphere. This is a model artifact and \ce{C2H} should be regarded as the precursor for more complex organics. It has been known that \ce{C2H} can be attached to existing hydrocarbons to form higher-order hydrocarbons and therefore hazes (e.g., Yung et al. 1984). We identify this caveat as an important aspect of further studies.

The scarcity of hydrogen is the main driver for the formation of unsaturated hydrocarbon in oxygen-poor carbon-rich thick atmospheres of super Earths and mini Neptunes. We systemize the speciation of carbon in hydrocarbon-rich atmospheres in Figure \ref{HYDROCARBON_T}, for which we have used thermochemistry simulations (i.e., minimizing global Gibbs free energy) to compute the relative abundance of hydrocarbons for different hydrogen abundances $X_\ce{H}$ and temperatures and pressures. For an intermediate $X_\ce{H}$, significant, even dominant amounts of \ce{C2H4} and \ce{C2H2} would be present in the atmosphere. For smaller and smaller $X_\ce{H}$, the hydrocarbons in the atmosphere become less and less saturated (Figure \ref{HYDROCARBON_T}). Interestingly, \ce{C2H6} cannot be the dominant carbon species in any cases. This is in contrast with the photochemical formation of unsaturated hydrocarbons (mostly in \ce{C2H6}) in Jupiter's atmosphere (e.g., Gladstone et al. 1996) and Titan's atmosphere (e.g., Yung et al. 1984), which is driven by the photon-initiated dissociation of methane and is confined to pressures lower than 0.1 bar (i.e., the stratosphere). Here, the formation of unsaturated hydrocarbon is no longer a photochemical perturbation, but an inevitable result of hydrogen loss of the atmosphere. Unsaturated hydrocarbons can be the dominant gases in evolved carbon-rich atmospheres on super Earths, and they lead to spectral features that allow unique identification of carbon-rich atmospheres.

The mixing ratios of \ce{CH4}, \ce{C2H2}, and \ce{C2H4} in the hydrocarbon-rich atmospheres also depend on the internal heat flux and the efficiency of vertical mixing. First, the internal heat flux can directly affects the abundance of hydrocarbons through controlling the temperature at the quenching pressure. The lower panel of Figure \ref{CHEM_INT} shows a case of $X_\ce{H}=0.5$ and $X_\ce{C}/X_\ce{O}=2$. This atmosphere is mainly composed of \ce{CO} and \ce{CH4}, but also has a great deal of \ce{C2H2} and \ce{C2H4}. Increasing $T_{\rm int}$ from 35 K to 180 K, the mixing ratio of \ce{C2H2} increases from $5\times10^{-3}$ to $5\times10^{-2}$, and the mixing ratio of \ce{C2H4} decreases from $10^{-1}$ to $5\times10^{-2}$. While the temperature changes dramatically by thousands of K at $10^3$ Bar when $T_{\rm int}$ increases from 35 K to 180 K (i.e., the planet's internal heat flux increases by 700 folds), the effect on the mixing ratios of the major components at the observable pressure levels is rather small. This is because elevating the internal heat flux moves up the quenching level and diminishes the effect of internal heating. Second, the internal heat flux affects the manner that the mixing ratios of hydrocarbons depend on the eddy diffusivity. When the eddy mixing is more efficient, the quenching pressure becomes higher. If the planet has negligible internal heat flux, the atmosphere below the radiative layer would be close to isothermal, and then as a result of increasing pressure and constant temperature, a greater eddy mixing coefficient would lead to more \ce{CH4} in the observable part of the atmosphere. If the planet has significant internal heat flux, the atmospheric temperature would increase with pressure adiabatically at depth, and then the mixing ratio of \ce{C2H2} should increase with more efficient vertical mixing (Figure \ref{HYDROCARBON_T}; also see the upper-right panel of Figure \ref{CHEM_KZZ}). In all, \ce{CH4}, \ce{C2H2}, and \ce{C2H4} have distinctive spectral features; and their relative abundance may provide a probe to the processes of eddy mixing and internal heating in the deep atmosphere of a super Earth that are otherwise not detectable.

The results above are valid for hydrocarbon-rich atmospheres, which typically requires $X_\ce{C}/X_\ce{O}\geq2$. If we loosen this requirement and consider the cases with $X_\ce{C}/X_\ce{O}\geq1$, we find that the atmosphere is most likely to be dominated in \ce{CO}. For  $X_\ce{C}/X_\ce{O}\sim1$, the abundance of unsaturated hydrocarbons in the non-\ce{H2}-dominated thick atmosphere of an exoplanet having similar temperatures as HD~97658~b would be at the level of 1000 ppmv in the observable part of the atmosphere, also significant in terms of driving spectral features. More efficient photolysis and less efficient vertical mixing would favor formation and accumulation of hydrocarbons. In all, we find that unsaturated hydrocarbons are abundant in non-\ce{H2}-dominated carbon-rich thick atmospheres on super Earths and mini Neptunes, and therefore they should be considered as one of the basic building blocks for super Earth/mini Neptune atmospheres.

Finally we turn to the minor components in hydrocarbon-rich atmospheres. The main form of oxygen in hydrocarbon-rich atmospheres is invariably \ce{CO}. For a non-\ce{H2}-dominated carbon-rich oxygen-poor atmosphere (i.e., $X_\ce{H}\leq0.7$, $X_\ce{C}/X_\ce{O}\geq2$), we always find that $X_\ce{H2O}/X_\ce{CO}<0.1$, and $X_\ce{H2O}/X_\ce{CO}$ becomes even smaller for higher temperatures or higher $X_\ce{C}/X_\ce{O}$ (Figure \ref{General_CHO} and Figure \ref{CHO_RATIO}).

\section{Application to Observations}

We here present synthesized spectra based on the photochemistry-thermochemistry simulations in order to compare with current observations and to guide future observations. We apply our self-consistent atmosphere models to known transiting low-mass exoplanet GJ~1214~b, HD~97658~b, and 55~Cnc~e. The results presented below are not only applicable to these planets, but also applicable to super Earths and mini Neptunes that receive similar degree of heating from their parent stars.

\subsection{{\rm  GJ~1214~b}}

\begin{figure}[h]
\begin{center}
 \includegraphics[width=0.7\textwidth]{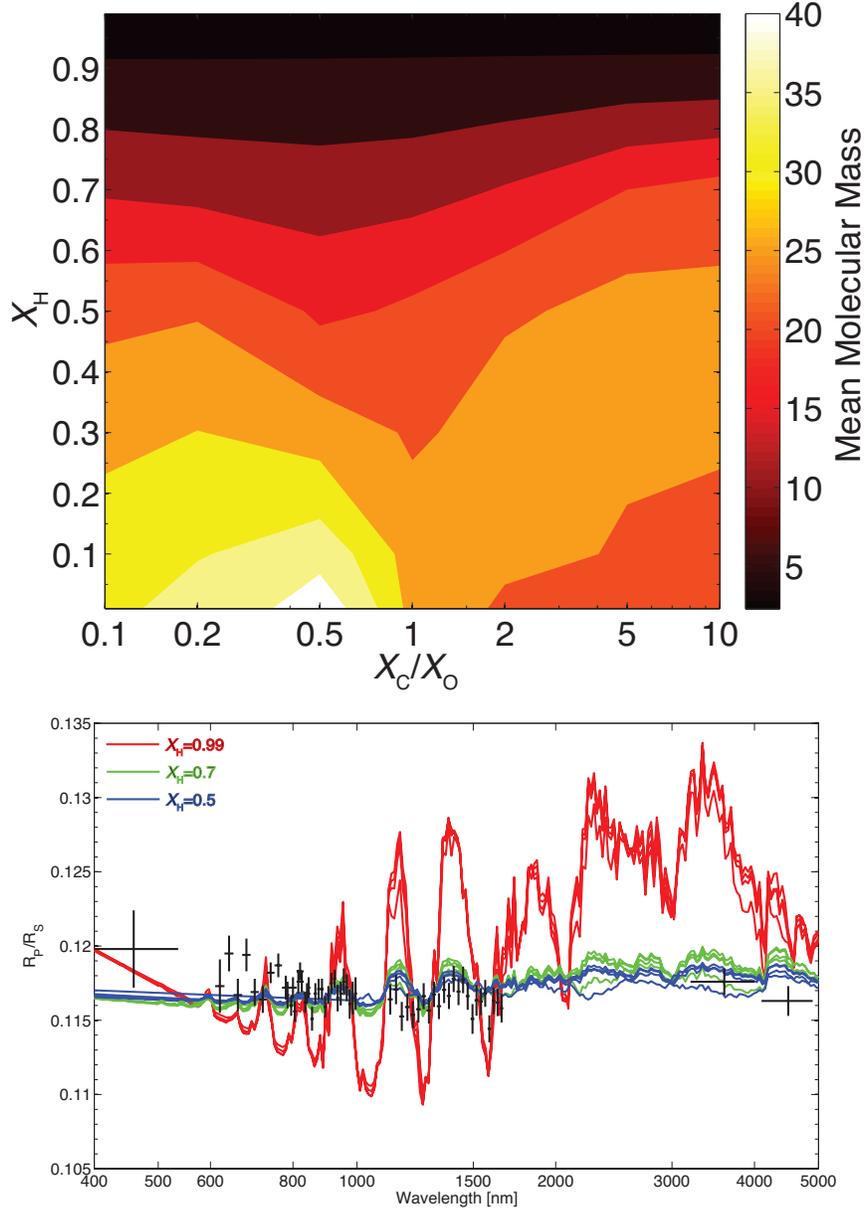}
 \caption{
 {\it Upper panel} The mean molecular mass of the observable part (1-100 mbar) of the modeled atmospheres of super Earth/mini Neptune GJ 1214 b. 
 {\it Lower panel} Modeled transmission spectra of GJ~1214~b in comparison with current observations.  The synthetic spectra are computed based on the results of the photochemistry-thermochemistry simulations for $X_\ce{H}=0.99$ (red lines), 0.7 (green lines), and 0.5 (blue lines) assuming an eddy diffusivity of $10^9$ cm$^2$ s$^{-1}$. For each $X_\ce{H}$, we show the spectra corresponding to the carbon to oxygen ratio of 0.1, 0.5, 1, and 2. Also plotted in the figure are the current observation data spanning from visible to mid-infrared wavelengths, from de Mooij et al. (2012), Bean et al. (2011), Berta et al. (2012), and D\'esert et al. (2011). The main point is that as long as $X_\ce{H}\leq0.7$, the mean molecular mass in the atmosphere would be greater than 15, regardless of the carbon to oxygen ratio, and the corresponding transmission spectrum would be flat enough to be consistent with current observations.
 }
 \label{SPEC_FIT}
  \end{center}
\end{figure}

Current observations of the super Earth/mini Neptune GJ~1214~b have shown a flat spectrum in transmission (Bean et al., 2010, 2011; Crossfield et al. 2011; D\'esert et al. 2011; Berta et al. 2012; de Mooij et al. 2012). The observed flat transmission spectrum has ruled out the scenario that the planet has a \ce{H2}-dominated clear atmosphere, but can be explained by an atmosphere having a mean molecular mass larger than 15, or an atmosphere having thick layers of haze (e.g., Bean et al. 2010; Miller-Ricci Kempton et al. 2012; Howe \& Burrows 2012; Benneke \& Seager 2012; Morley et al. 2013). Using the photochemistry-thermochemistry model and focusing on atmospheric scenarios without haze or clouds, we find that as long as $X_\ce{H}\leq0.7$, the mean molecular mass of the observable part of the atmosphere is greater than about 15, and such an atmosphere would be consistent with current observations (Figure \ref{SPEC_FIT}). All models with $X_\ce{H}\leq0.7$, regardless of the carbon to oxygen ratio, provide adequate fit to the current observations of GJ~1214~b (Figure \ref{SPEC_FIT}). The acceptable scenarios based on the transmission spectrum includes water-dominated atmosphere (upper-left panel of Figure \ref{CHEM_KZZ}), \ce{H2}-\ce{CO} dominated atmosphere (lower-left panel of Figure \ref{CHEM_KZZ}), \ce{CO}-\ce{CH4} dominated atmosphere (upper-right panel of Figure \ref{CHEM_KZZ}), and \ce{C2H2}-\ce{C2H4} dominated atmosphere (lower-right panel of Figure \ref{CHEM_KZZ}).

%\begin{figure}[h]
%\begin{center}
% \includegraphics[width=0.45\textwidth]{GJ1214_MeanMolecular.eps}
% \caption{
% Mean molecular mass of the visible part (pressure ranging from 1 mbar to 1 bar) of the atmosphere on GJ 1214 b as depended on the elemental abundance. The mean molecular mass is synthesized based on the photochemical simulations shown in Figure \ref{Zoo}. 
% As long as $H\leq0.7$, the mean molecular mass in the atmosphere would be larger than 15, thus consistent with current observations.
%  }
% \label{MeanMolecular}
%  \end{center}
%\end{figure}

% A large number of atmospheric scenarios are consistent with the current observations of GJ~1214~b. Depending on the carbon to oxygen ratio of the atmosphere, plausible scenarios include: water-dominated atmosphere (upper panel of Figure \ref{CHEM_KZZ}), \ce{H2}-\ce{CO} dominated atmosphere (lower panel of Figure \ref{CHEM_KZZ}), \ce{CO}-\ce{CH4} dominated atmosphere (upper panel of Figure \ref{CHEM_KZZ}), and \ce{C2H2}-\ce{C2H4} dominated atmosphere (lower panel of Figure \ref{CHEM_KZZ}). The main controlling factor that determines the composition is the carbon to oxygen ratio, while vertical mixing may have an effect in certain scenarios. The list of plausible scenarios may be completed by including the scenarios of hydrogen-rich atmospheres with high-altitude hazes that are not treated in our models (Bean et al. 2010; Miller-Ricci Kempton et al. 2012; Howe \& Burrows 2012).

\begin{figure*}[h]
\begin{center}
 \includegraphics[width=0.95\textwidth]{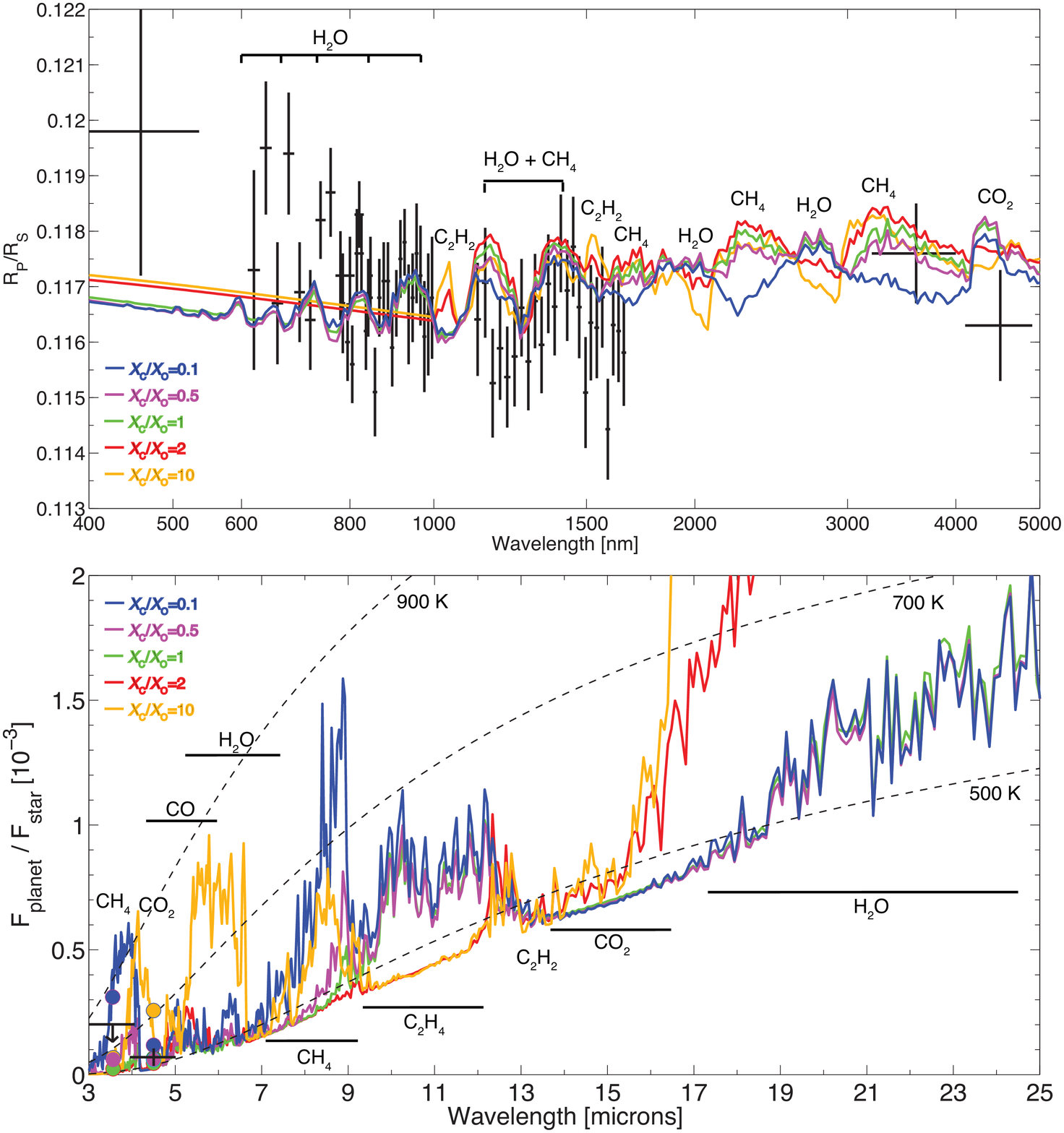}
 \caption{
 Molecular features in the transmission spectra and thermal emission spectra of non-\ce{H2}-dominated atmospheres on GJ~1214~b based on our photochemistry-thermochemistry simulations. The simulated spectra are for $X_\ce{H}=0.5$ and a variety of carbon to oxygen ratios ranging from oxygen rich to carbon rich, and for an eddy diffusivity of $10^9$ cm$^2$ s$^{-1}$. Also plotted in the figure are the current observation data in transmission spanning from visible to mid-infrared wavelengths (references listed in the caption of Figure \ref{SPEC_FIT}), and the latest thermal emission constraints in the {\it Spitzer} bands (Gillon et al. 2013) in comparison with model predicted emission fluxes with the {\it Spitzer} bandpass incorporated (filled circles). The atmospheric scenarios with different carbon to oxygen ratios can be constrained via the spectral features of their hallmark molecules (see the highlights of the available spectral features in Section 4.1). }
 \label{SPEC_FEATURE}
  \end{center}
\end{figure*}

How could future observations distinguish these scenarios? Figure \ref{SPEC_FEATURE} shows the transmission spectra and the thermal emission spectra of the planet if its atmosphere in non-\ce{H2}-dominated. A number of diagnostic features indicating hallmark molecules for different atmospheric scenarios stand out, and these features will allow future characterization of super Earth/mini Neptune GJ~1214~b, and other transiting super Earths and mini Neptunes with similar temperatures to be discovered. Out of a number of spectral features labeled in Figure \ref{SPEC_FEATURE}, we highlight the following three points.

First, a hydrocarbon-rich atmosphere (i.e., the case of $X_\ce{C}/X_\ce{O}\ge2$) can be uniquely identified by detecting the absorption bands of \ce{C2H2} at 1.0 and 1.5 $\mu$m in transmission and the absorption bands of \ce{C2H2} and \ce{C2H4} at 9 - 14 $\mu$m in thermal emission. In particular, the \ce{C2H2} feature at 1.0 $\mu$m is not contaminated by other potential constituents in the atmosphere, and is as sizable as the water features nearby (Figure \ref{SPEC_FEATURE}). The photochemistry-thermochemistry models show that \ce{C2H2} is expected to be abundant in a carbon-rich atmosphere of GJ 1214, and we suggest \ce{C2H2} as a main component and hallmark molecule for carbon-rich atmospheres on a GJ~1214~b like exoplanet. To complicate the matter, transmission spectroscopy is often interfered by haze in the atmosphere (Fortney 2005; Howe \& Burrows 2012; Benneke \& Seager 2012; Morley et al. 2013). Although we do not treat photochemical haze in this paper, we expect that non-\ce{H2}-dominated carbon-rich atmospheres to be ideal environment to form photochemical haze, and therefore the transmission spectra produced by such kind of atmospheres are likely to be affected by haze.

Second, measurements of the thermal emission of this planet in the mid-infrared wavelengths yields a great deal of knowledge of its atmosphere. A hydrocarbon-rich atmosphere has little \ce{H2O} or \ce{CO2}, so its thermal emission spectrum should be dominated by the absorption bands of \ce{CH4}, \ce{C2H4}, and \ce{C2H2} (Figure \ref{SPEC_FEATURE}). A water-rich atmosphere has little methane or other hydrocarbons,  so its thermal emission spectrum should only show the prominent absorption bands of \ce{CO2} on top of the pseudo-continuum of \ce{H2O} absorption (Figure \ref{SPEC_FEATURE}). Recent observations with {\it Spitzer} have constrained the planet's thermal emission at 4.5 $\mu$m to be $70\pm35$ ppm of stellar emission and at 3.6 $\mu$m to be not greater than 205 ppm. Figure \ref{SPEC_FEATURE} shows that these constraints rule out the cloud-free scenarios with a very high or very low carbon to oxygen ratio ($X_\ce{C}/X_\ce{O}=0.1$ or $X_\ce{C}/X_\ce{O}=10$), and permits all other scenarios with $X_\ce{C}/X_\ce{O}$ ranging from 0.5 to 2. The thermal emission spectrum is likely to be observed by future infrared telescopes in space, as the planet to star flux ratio will be within the reach of the {\it JWST}. 

Third, a water-dominated atmosphere may be inferred from the nonexistence of methane absorption features. With transmission and thermal emission spectroscopy, a water-rich atmosphere (i.e., the case of $X_\ce{C}/X_\ce{O}\leq0.5$) may be inferred based on the detection of water vapor absorption bands in the near-infrared. However, it would be very hard to establish that the atmosphere is mostly composed of water vapor based on water vapor features alone. Figure \ref{SPEC_FEATURE} shows that the atmosphere scenarios with $X_\ce{C}/X_\ce{O}$ ranging from 0.1 to 1, which have the water vapor mixing ratio ranging from 70\% to 5\%, have almost identical water vapor features in transmission at 0.6-1.5 $\mu$m and in thermal emission. The water vapor features would allow detection of {\it water-rich} atmospheres but not {\it water-dominated} atmospheres. The photochemistry-thermochemistry models show that a \ce{H2}-depleted water-dominated atmosphere must have \ce{CO2} as the carbon carrier and must not have any significant amounts of \ce{CH4}. Therefore, a confirmed nonexistence of methane feature in either transmission or thermal emission (see Figure \ref{SPEC_FEATURE}), in combination with a detection of water features, will provide sufficient evidence for a water-dominated atmosphere on a GJ~1214~b like super Earth.

% Interestingly, both water and methane absorbs in the 1-2 $\mu$m band; therefore this band is not favorable to distinguish a water rich atmosphere versus a hydrocarbon rich atmosphere. 

% The intermediate cases, for which the dominant gas in the atmosphere is \ce{CO} (Figure \ref{CHO}), and water and methane has equally trace amounts, may be inferred from the coexistence of water feature and methane feature in the transmission spectra.

\subsection{{\rm HD~97658~b}}

\begin{figure*}[h]
\begin{center}
 \includegraphics[width=1.0\textwidth]{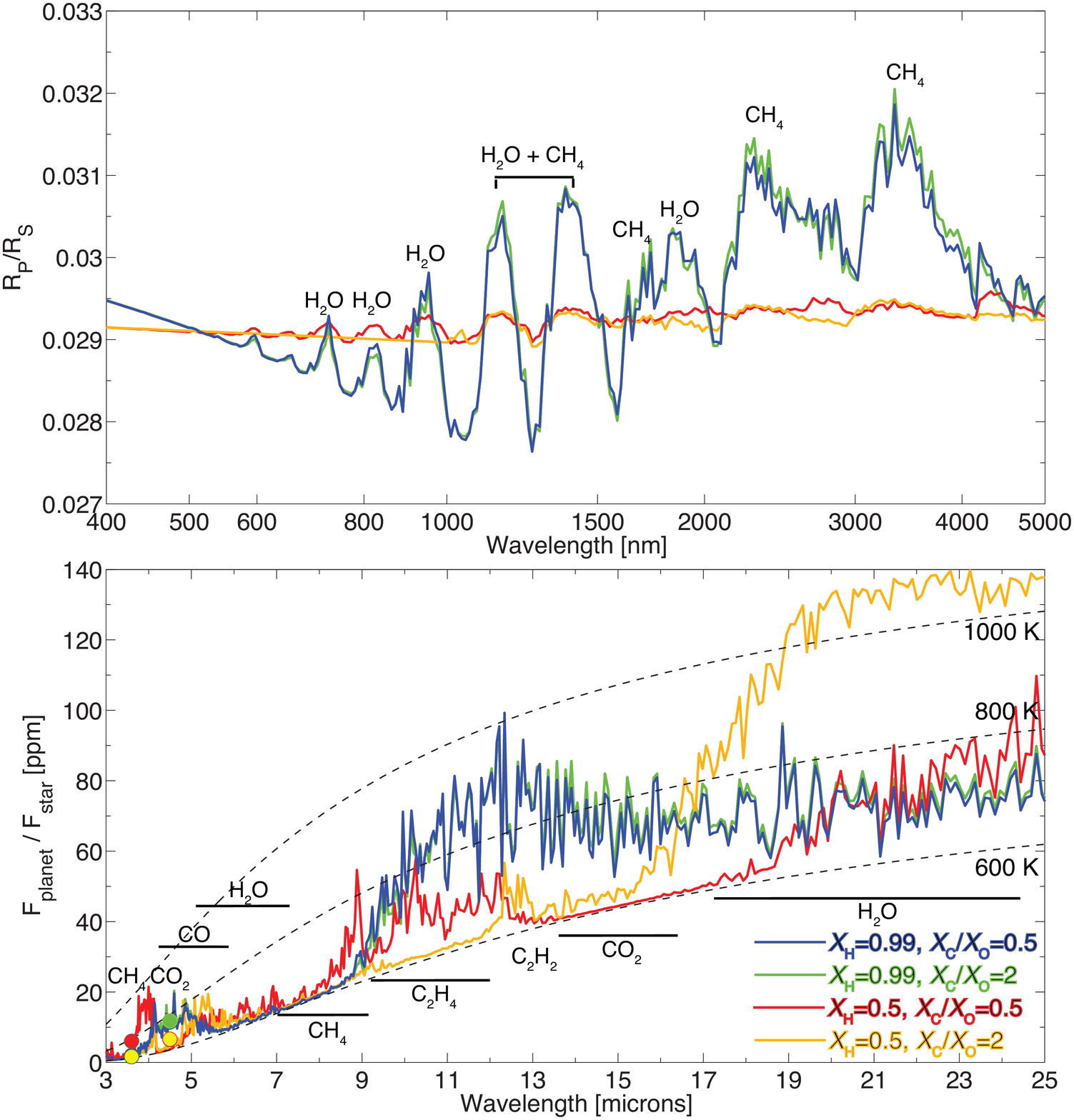}
 \caption{
 Modeled transmission spectra and thermal emission spectra for transiting super Earth/mini Neptune HD~97658~b based on our photochemistry-thermochemistry simulations. The presented spectra are for both \ce{H2}-dominated ($X_\ce{H}=0.99$) and non-\ce{H2}-dominated ($X_\ce{H}=0.5$) atmospheres with a variety of carbon to oxygen ratios ranging from oxygen-rich to carbon-rich, and for an eddy diffusivity of $10^9$ cm$^2$ s$^{-1}$. The filled circles are the model predicted planetary emission fluxes at 3.6 and 4.5 $\mu$m with the {\it Spitzer} bandpass incorporated. Performing near-infrared transmission spectroscopy for this system should find combined features of water and methane for a wide range of carbon to oxygen ratios. The thermal emission spectrum contains prominent molecular features that can differentiate an oxygen-rich atmosphere versus a carbon-rich atmosphere when the atmosphere is non-\ce{H2}-dominated.}
 \label{SPEC_FEATURE_HD97658B}
  \end{center}
\end{figure*}

HD~97658~b is a $2.34^{+0.18}_{-0.15}$ R$_{\oplus}$ super Earth/mini Neptune discovered recently (Dragomir et al. 2013). Orbiting a very bright star, the transiting super Earth/mini Neptune is one of the prime target for follow-up observations to determine its atmospheric properties in the near future. The planet receives stellar radiation flux of $T_{\rm irr}=1030$ K, making it warmer than GJ~1214~b and cooler than 55~Cnc~e. The measured mass and radius of the planet implies that the planet must have a significant gas envelope which can be either \ce{H2}-dominated or non-\ce{H2}-dominated (Dragomir et al. 2013). Using our photochemistry-thermochemistry models, we compute self-consistent atmosphere models for this planet and show synthetic spectra in Figure \ref{SPEC_FEATURE_HD97658B}.

The transmission spectrum of HD~97658~b will be dominated by the combined feature of water and methane in the near-infrared wavelengths if the planet has an \ce{H2}-dominated cloud-free atmosphere. For the temperature of the planet, methane should be the dominant form of carbon and water should be the dominant form of oxygen in the \ce{H2}-dominated atmosphere. Since the mixing ratios of \ce{CH4} and \ce{H2O} only have linear dependency on $X_\ce{C}/X_\ce{O}$, it will be difficult to distinguish a carbon-rich atmosphere versus a oxygen-rich atmosphere. 

If the planet has a non-\ce{H2}-dominated atmosphere, the size of the features in its transmission spectrum will be quite small, but its thermal emission spectrum is expected to have prominent features of \ce{CO2} for an oxygen-rich atmosphere or \ce{C2H2} and \ce{C2H4} for a carbon-rich atmosphere (Figure \ref{SPEC_FEATURE_HD97658B}). Therefore a carbon-rich atmosphere and a oxygen-rich atmosphere can be distinguished by thermal emission spectroscopy in the future if transmission spectroscopy finds the planet's atmosphere to be non-\ce{H2}-dominated. In both warm {\it Spitzer} bands, the planet's flux is within 10 ppm of the stellar flux, making the planet's emission very difficult to be observed by {\it Spitzer}. In the future it is promising to perform detail atmosphere characterization for this planet by mid-infrared spectroscopy with {\it JWST} because different atmosphere scenarios (\ce{H2}-dominated versus non-\ce{H2}-dominated, carbon-rich versus oxygen-rich) have distinctive molecular features (see Figure \ref{SPEC_FEATURE_HD97658B}).

\subsection{{\rm 55~Cnc~e}}

The measured radius of super Earth 55~Cnc~e ($2.2\pm0.1$ R$_{\oplus}$) implies a volatile envelope on the planet (Gillon et al. 2012). For the measured mass of the planet, the planet's radius would be 1.5 $R_{\oplus}$ if it was completely composed of iron, or 1.9 $R_{\oplus}$ if it was completely composed of \ce{MgSiO3} (estimated using of the models of Zeng \& Sasselov 2013); that is to say, the room for a potential atmosphere is maximally $\sim4000$ km thick, or 30\% of the observed planet radius. The planet's radius has also been measured by Winn et al. (2011) in the visible wavelengths to be $\sim2$ $R_{\oplus}$ (based on reanalysis by Gillon et al. 2012). Recently, the detection of the thermal emission of the planet in the {\it Spitzer} 4.5-$\mu$m band indicates a relatively high dayside brightness temperature ($2360\pm300$ K; Demory et al. 2012).

\begin{figure*}[h]
\begin{center}
 \includegraphics[width=0.95\textwidth]{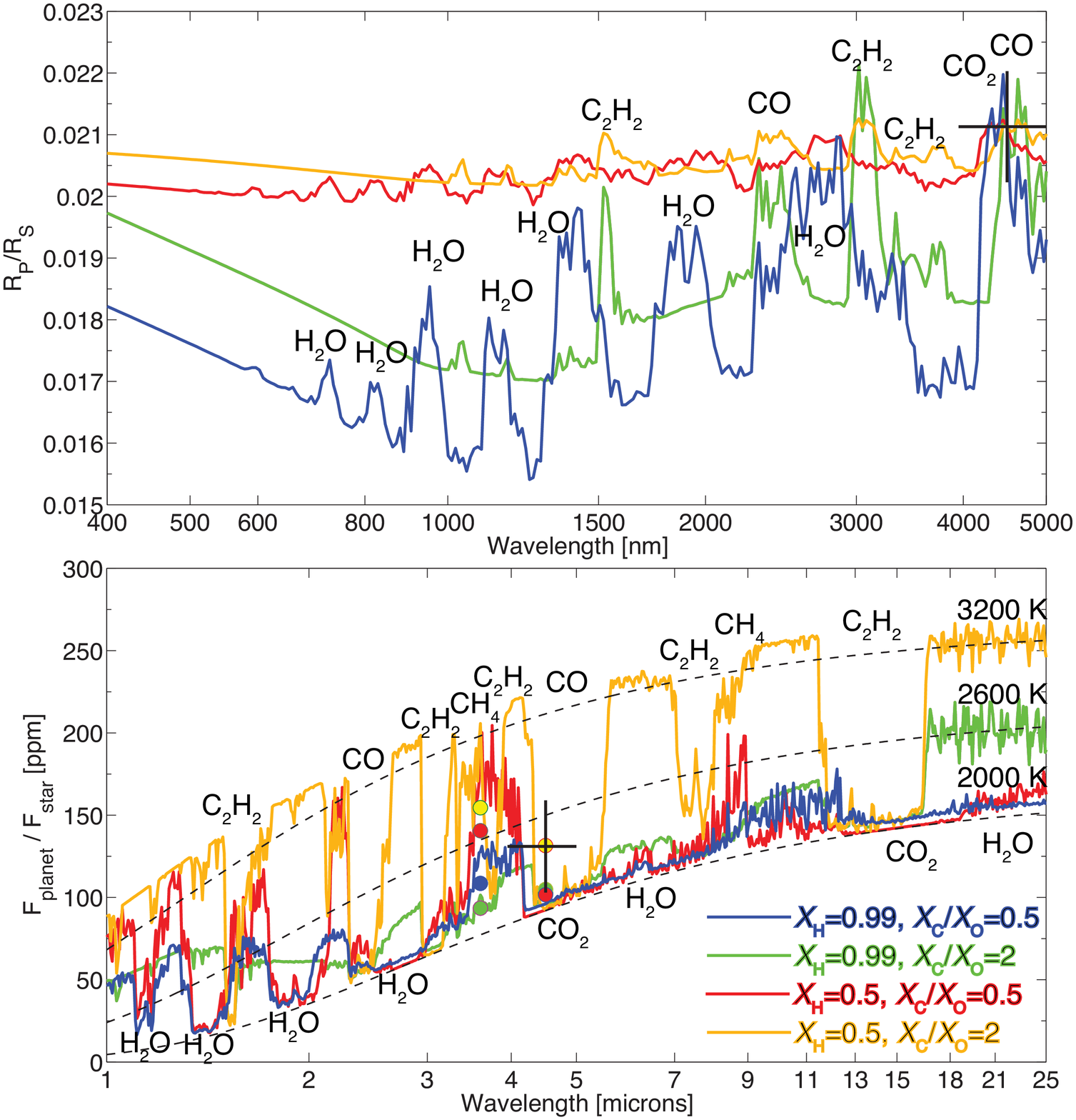}
 \caption{
 Modeled transmission spectra and thermal emission spectra of thick atmospheres on a 55~Cnc~e like super Earth based on photochemistry-thermochemistry simulations. The presented spectra are for both \ce{H2}-dominated ($X_\ce{H}=0.99$) and non-\ce{H2}-dominated ($X_\ce{H}=0.5$) atmospheres with a variety of carbon to oxygen ratios ranging from oxygen-rich to carbon-rich, and for an eddy diffusivity of $10^9$ cm$^2$ s$^{-1}$. Overplotted in the figure are the current observation data in transmission (Gillon et al. 2012) and in thermal emission (Demory et al. 2012). The filled circles are the model predicted planetary emission fluxes at 3.6 and 4.5 $\mu$m with the {\it Spitzer} bandpass incorporated. The cases of $X_\ce{H}=0.99$ would necessarily require the planet's interior to be mainly composed of iron, in order to satisfy the mass-radius constraints.
The diagnostic features of water-rich atmospheres (\ce{H2O} and \ce{CO2} features, labeled below the spectra) and the diagnostic features of hydrocarbon-rich atmospheres (\ce{C2H2} and \ce{CO} features, labeled above the spectra) are well separated in both transmission and thermal emission spectrum.
% Note that our models may under-predict the apparent size of the planet in the visible wavelengths for the cases of hydrogen-rich atmospheres, because there could be strong absorbers in the visible wavelengths in the conditions of 55~Cnc~e that are not considered in our model (such as TiO, VO, and condensate particles).  
 }
 \label{SPEC_FEATURE_55CNCE}
  \end{center}
\end{figure*}

We simulate the transmission spectra and thermal emission spectra of a 55~Cnc~e like super Earth, assuming a thick atmosphere on the planet (Figure \ref{SPEC_FEATURE_55CNCE}).  Current observations of the thermal emission of 55~Cnc~e in the {\it Spitzer} 4.5-$\mu$m band is likely to contain absorption features of either \ce{CO2} in oxygen-rich scenarios, or \ce{CO} in carbon-rich scenarios, if the planet has an atmosphere. Assuming poor heat redistribution, all our atmosphere models are consistent with the {\it Spitzer} 4.5-$\mu$m observation at 1-$\sigma$. Interestingly, the {\it Spitzer} 3.6-$\mu$m band (not yet observed), where the main \ce{CH4} band falls into, is likely to probe deep into the atmosphere and has a higher brightness temperature than the 4.5-$\mu$m band, because methane is not expected to exist in significant amounts in such a high-temperature atmosphere according to our photochemistry-thermochemistry simulations.

For future observations of 55~Cnc~e, a water-rich atmosphere can be detected in thermal emission with the absorption bands of \ce{H2O} and \ce{CO2}, and by transmission spectroscopy if the mean molecular mass is low (Figure \ref{SPEC_FEATURE_55CNCE}). This is similar to the case of GJ~1214~b that has a much lower temperature. A hydrocarbon-rich atmosphere on 55~Cnc~e would result in prominent absorption bands of \ce{CO} and \ce{C2H2}, potentially detectable via thermal emission, and transmission if the mean molecular mass is low (Figure \ref{SPEC_FEATURE_55CNCE}). Note that unlike the case of GJ~1214~b, the nonexistence of methane feature cannot be interpreted as the water dominance. On 55~Cnc~e, the methane features are expected to be suppressed due to the high temperature.

The transmission spectroscopy of 55~Cnc~e, if the planet has an extended \ce{H2}-rich atmosphere, is within reach of current observation facilities including {\it VLT} and {\it Hubble}. Especially, the diagnostic features of water-rich atmospheres (\ce{H2O} features) and the diagnostic features of hydrocarbon-rich atmospheres (\ce{C2H2} and \ce{CO} features) are well separated in the near-infrared wavelengths in transmission spectra (Figure \ref{SPEC_FEATURE_55CNCE}), which would enable straightforward characterization of the chemical compositions of the atmosphere.

\section{Discussion}

%\subsection{Water Worlds Could be Rare}

\subsection{Finding Water-Dominated Thick Atmospheres May be Challenging}

Water-dominated atmosphere, if discovered, would be an exciting environment in our interstellar neighborhood because of its reminiscence of a habitable world. In fact, one of preferred scenarios for the super Earth/mini Neptune GJ~1214~b is that the planet has a signficant fraction of its mass as water based on mass-radius constraints (Rogers \& Seager 2010b; Nettelmann et al. 2011) and that the planet has a water-dominated atmosphere or a \ce{H2}-\ce{H2O} atmosphere (Nettelmann et al. 2011) based on the flat transmission spectrum (Bean et al. 2010, 2011; Berta et al. 2012). We study the range of atmospheric composition of GJ~1214~b in detail using our photochemistry-thermochemistry, and find that the water-dominated atmosphere exists at a carbon to oxygen ratio much smaller than the solar value. For a range of carbon to oxygen ratios, we show a range of possible scenarios and their transmission and thermal emission spectra, and outline a number of possible ways to distinguish them by future observations in Section 4.1.

For a more general discussion beyond GJ~1214~b, we suggest that identifying a water-dominated atmosphere may be harder than previously expected. There are two reasons for this suggestion. The first reason is that water vapor features in the transmission spectra are not effective in distinguishing a water-dominated atmosphere versus a water-rich atmosphere. This has been shown by Benneke \& Seager (2012) with a quantitative retrieval method for super Earth transmission spectra. This is also shown in Figure \ref{SPEC_FEATURE} as an example, in which atmospheric scenarios with water vapor mixing ratios ranging from 1\% to 70\% have similar sizes of water vapor features in the transmission spectra. The second reason is that the mixing ratio of water vapor in a non-\ce{H2}-dominated atmosphere is highly sensitive to the carbon to oxygen ratio (Figure \ref{CHO_RATIO}). In particular, for a solar carbon to oxygen ratio, a non-\ce{H2}-dominated atmosphere is not water-dominated, but such an atmosphere will have prominent water vapor features in its spectrum (Figure \ref{CHEM_KZZ} and Figure \ref{SPEC_FEATURE}). Therefore, identifying a water-dominated atmosphere not only involves measuring the water vapor features, but also require evaluating the carbon to oxygen ratio of the atmosphere, which warrants measurements of other gases in the atmosphere.

Moreover, a water-dominated atmosphere might be indigenously rare in nature. For a planet to have a water-dominated atmosphere, the planet has to start with an almost pure water ice envelope, with the fraction of methane ice of no more than 20\% by weight. We show in section 3.2 that the carbon to oxygen ratio has to be less than 0.2, much smaller than the solar ratio, for a thick atmosphere to evolve into a water-dominated atmosphere. Such a condition on the planet's forming environment is quite confining, which might only be possible close to the snow line of a protoplanetary disk with a solar-like carbon to oxygen ratio. Studies of the condensation sequence of volatile ices from protoplanetary nebula gases have shown that the ice compositions sensitively depend on the carbon to oxygen ratio of the nebulae (e.g., Marboeuf et al. 2008; Johnson et al. 2012). Even for a planet-forming nabula with a solar carbon to oxygen ratio, the ice mixture should have $\sim20$\% carbon by weight (Marboeuf et al. 2008); and the carbon content in ice increases with increasing carbon to oxygen ratios of the nebulae, in particular for those with cold midplanes (Johnson et al. 2012). To summarize, one could expect the super Earths that are born as mini Neptunes to have a wide range of carbon to oxygen ratios in their atmospheres, and a water-dominated atmosphere is one of many plausible atmospheric scenarios. This finding suggests that water-dominated atmospheres might be rare.

\subsection{Chemical Stability of Atmospheric Gases}

One useful application of our photochemistry-thermochemistry model is to verify the chemical stability of atmospheric gases. The ``million model" approach has been developed to interpret the observations of exoplanet atmospheres (e.g., Madhusudhan \& Seager 2009; Benneke \& Seager 2012). These methods seek best-fitted model spectra of planetary emission and transmission by exploring the full parameter space of molecular compositions. We have shown in this paper that not all combinations of gases are chemically stable in thick atmospheres on super Earths and mini Neptunes. For example, we find that methane is not compatible with water-dominated atmospheres on GJ~1214~b. Therefore, our photochemistry-thermochemistry model can be used in conjunction with the ``million model'' approach to eliminate a significant part of the parameter space, in terms of ruling out the unstable atmospheric scenarios and strengthening the interpretation of spectra.

As a general study on the stability of gases in thick atmospheres, we additionally simulate atmospheres initially composed of all possible 50\%-50\% molecular binary combination of \ce{H2}, \ce{H2O}, \ce{CH4}, \ce{CO}, and \ce{CO2} for an exoplanet like GJ~1214~b. For each scenario, we derive the elemental abundance from the initial composition for the model input. For example, to simulate a 50\% \ce{H2} 50\% \ce{CO} atmosphere, we use an H:O:C ratio of 2:1:1. The thermochemistry-photochemistry model result is then compared with the initial molecular composition, to determine whether such an initial state is chemically stable.

\begin{figure}
\begin{center}
 \includegraphics[width=0.7\textwidth]{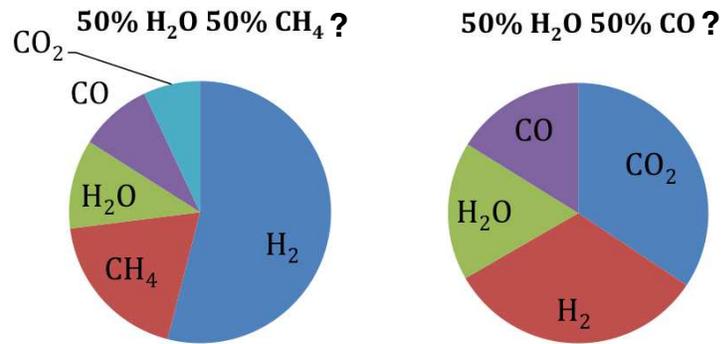}
 \caption[Compatibility of gases in thick atmospheres on terrestrial exoplanets]{Compatibility of gases in thick atmospheres on terrestrial exoplanets. Mixing ratios of main components in the observable part of the atmosphere on an exoplanet like GJ~1214~b, based on photochemistry-thermochemistry simulations starting from an initial composition of 50\% \ce{H2O} and 50\% \ce{CH4}, and an initial composition of 50\% \ce{H2O} and 50\% \ce{CO}. The initial compositions are not chemically stable because \ce{H2O} will oxidize \ce{CH4} or \ce{CO} in the mixture. Therefore 50\% \ce{H2O}-50\% \ce{CH4}, or 50\% \ce{H2O}-50\% \ce{CO}, are not plausible scenarios for the atmosphere on GJ~1214~b.
  }
 \label{Compa}
  \end{center}
\end{figure}

We find that \ce{H2} is compatible with all major stable gases of C, H, O elements, but \ce{H2O} is not compatible with equal amount of \ce{CH4} and \ce{CO}. An atmosphere mainly composed of \ce{H2O} and \ce{CO} or \ce{CH4} is not stable for GJ~1214~b because of the following reactions:
\reaction{H2O + CH4 -> 3H2 + CO .}
\reaction{H2O + CO -> H2 + CO2 .}
For the elemental abundance of \ce{H2O}-\ce{CO} or \ce{H2O}-\ce{CH4} combination, we find that \ce{CO} and \ce{CH4} can always reduce most of \ce{H2O} to \ce{H2} (see Figure \ref{Compa}). Similarly, the \ce{CH4}-\ce{CO2} combination has the same elemental abundance as the \ce{H2}-\ce{CO} combination; the atmosphere will have the composition of \ce{H2} and \ce{CO} at the steady state, and the \ce{CH4}-\ce{CO2} combination is not stable (and therefore not a plausible scenario). These results imply that chemical stability has to be taken into account when deriving atmospheric molecular compositions from spectra of super Earths and mini Neptunes.

\subsection{To What Extent do the Disequilibrium Processes Matter?}

One might ask whether a kinetic-transport simulation is necessary for the study of super Earths and mini Neptunes, considering the fact that observations of their atmospheres will remain disk-averaged and have low spectral resolution for a long time. Here, our two-level model framework (Figure \ref{ModelSchematic}) provides an opportunity to compare thermochemical equilibrium calculation with kinetic-transport calculation for a wide range of exoplanet thick atmospheres.

{\it Temperature profile} To what extent the temperature profile calculated under the assumption of thermochemical equilibrium would differ from the profile consistent with the chemical composition resulted from the kinetic-transport modeling? We find that among all models we have run, the variation of temperature caused by disequilibrium processes (i.e., photochemistry and vertical transport) is maximally 10\% with respect to the temperature profile that one would get assuming thermochemical equilibrium. In fact, for more than 90\% of the cases, the effect of disequilibrium chemistry on the temperature is within 5\%. For most cases the temperature would increases when considering disequilibrium processes like vertical transport, and the magnitude of such temperature variation increases with increasing eddy diffusivity. The most significant variation for the GJ~1214~b simulations happen for $X_{\rm H}=0.9$ and $0.5<X_{\rm C}/X_{\rm O}<2$, and $X_{\rm H}=0.7$ and $X_{\rm C}/X_{\rm O}=1$. In these scenarios, the increase of temperature is mostly due to \ce{H2O} and \ce{CH4} brought up by vertical transport to the pressure levels of 1-100 mbar.

\begin{figure}
\begin{center}
 \includegraphics[width=1.0\textwidth]{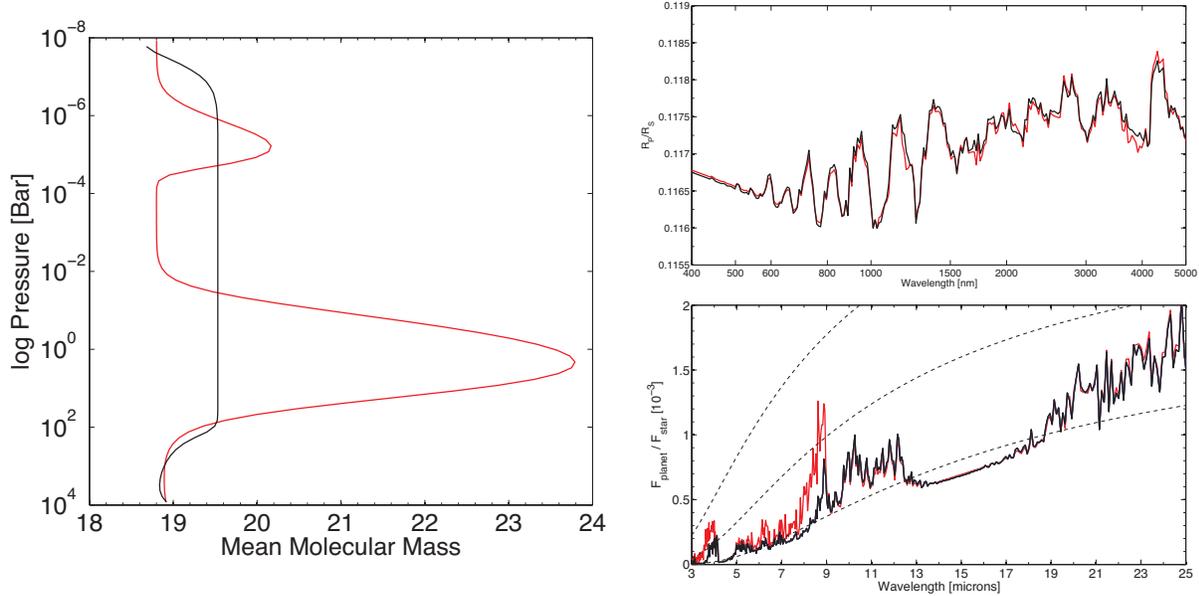}
 \caption[Comparison of the mean molecular mass and planetary spectra between the thermochemistry model and the kinetic-transport model]{Comparison of the mean molecular mass and planetary spectra between the thermochemistry model and the kinetic-transport model.
The simulation is for a thick atmosphere on a GJ~1214~b like exoplanet, having $X_{\rm H}=0.5$ and $X_{\rm C}/X_{\rm O}=0.5$. The red lines are from the thermochemical equilibrium model, and the black lines are from the kinetic-transport model with a vertical eddy diffusion coefficient of $10^9$ cm$^{2}$ s$^{-1}$.
The mean molecular mass ``bump" under thermochemical equilibrium at $\sim1$ Bar and $\sim10^{-5}$ Bar represents that at these pressures the main components of the atmosphere change from \ce{H2} and \ce{CO} to \ce{CO2} and \ce{CH4}; such change would not occur with vertical mixing. 
The thermochemical equilibrium model predicts a much higher emission flux in 7-9 $\mu$m than the model with vertical mixing, due to a lack of methane transported from below to 1-100 mbar.
This example shows that the kinetic-transport model that treats disequilibrium chemistry is critical to properly study the atmospheres on super Earths and mini Neptunes.
  }
 \label{MMZ_Var}
  \end{center}
\end{figure}

{\it Mean molecular mass and atmospheric structure} The same question can be asked for the mean molecular mass of the atmosphere. We find that the most significant variation in the mean molecular mass caused by disequilibrium processes is decrease of the mean molecular mass above the pressure level of $\sim0.1$ Pa due to lift of light species by molecular diffusion. Other than this, we also find that the fraction of major components of the atmosphere (and therefore the mean molecular mass) can vary significantly due to vertical transport only in the middle regime of Figure \ref{Zoo} where \ce{H2}, \ce{CH4}, \ce{CO}, \ce{CO2}, and \ce{H2O} can coexist. The main effect is that vertical transport may homogenize the atmosphere and remove the dependency of mean molecular mass on pressure under thermochemical equilibrium, when the temperature is not too high. One such example is shown in Figure \ref{MMZ_Var}. In this example, the difference between the true mean molecular mass and the mean molecular mass consistent from thermochemical equilibrium is up to 40\% in the observable part of the atmosphere. Similar effects, but to a lesser extent, occur for other scenarios in the parameter regime of $0.3<X_{\rm H}<0.7$ and $0.5<X_{\rm C}/X_{\rm O}<2$. This finding demonstrates the necessity of the kinetic-transport computation for a self-consistent modeling of thick atmospheres on warm super Earths and mini Neptunes.

{\it Planetary spectrum} The question that how disequilibrium chemistry would affect the emergent spectrum of a super Earth/mini Neptune is fairly involved, because the planetary spectrum depends on both the atmospheric thermal structure and the atmospheric chemical composition. We did not find a definitive trend comparing the spectra calculated for thermochemical equilibrium compositions and the spectra calculated for disequilibrium compositions. For transmission spectrum, our exploration of elemental abundances and temperatures shows that the error induced by assuming thermochemical equilibrium is in general within 1\% of the planet's radius, and this error is mostly due to mean molecular mass differences. For thermal emission spectrum, we find that in some cases disequilibrium processes can cause very significant variation, for which an example is shown in Figure \ref{MMZ_Var}. In this example the synthetic spectrum based on thermochemical equilibrium would significantly under-predict absorption of methane, which is expected to be enhanced by vertical transport. Therefore we suggest that self-consistent models are needed for understanding the physics and chemistry of thick atmospheres on super Earths and mini Neptunes.

\subsection{Trace Gases as the Probe for Vertical Mixing and Internal Heating}

Molecular compositions of a thick atmosphere at the pressure levels relevant to observations are controlled by vertical transport and chemical reactions at a deeper level. The idea of disequilibrium chemistry driven by vertical transport was originally proposed to explain the overabundance of \ce{CO} and the deficit of \ce{NH3} in Jupiter's atmosphere (Prinn \& Barshay 1977; Prinn \& Olaguer 1981) and used to explain chemical composition of Solar-System giant planets (e.g., Fegley \& Prinn 1985; Yung et al. 1988; Fegley \& Lodders 1994). Similar processes have also been suggested to operate on brown dwarfs and hot Jupiters (e.g., Fegley \& Lodders 1996; Griffith \& Yelle, 1999; Cooper \& Showman 2006; Line et al. 2010; Madhusudhan \& Seager 2011; Visscher \& Moses 2011; Moses et al. 2011). In general, the chemical timescale of a certain molecule decreases with increasing pressure and temperature. At the pressure levels where vertical transport timescale is shorter than the chemical kinetic timescale, the molecule is well mixed; at deeper levels in which the chemical timescale becomes shorter than the vertical transport timescale, the molecule's abundance is set by thermochemical equilibrium. Therefore, the molecule's abundance at observable pressure levels is set by thermochemical equilibrium at the level at which the vertical transport timescale equals the chemical kinetic timescale, i.e., the quenching level.

The most significant effects of vertical mixing are on the second most abundant carrier of major elements, and the most abundant carrier of minor elements. The abundances of the most abundant carriers of major elements, like \ce{H2}, \ce{H2O}, and \ce{O2} in water-rich atmospheres, are determined by the constraints of elemental abundances. Indeed, the elemental abundance could uniquely determine the abundances of major components in some super Earth thick atmospheres (see Table \ref{Para}). Based on the major composition set by the elemental abundance, relatively minor constituents in the atmosphere are indicators of the effects of eddy mixing and disequilibrium chemistry. The minor constituents can be either the carriers of minor elements, like carbon species in a water-rich atmosphere (Figure \ref{CHEM_KZZ}) and oxygen species in a hydrocarbon-rich atmosphere (Figure \ref{CHEM_KZZ}), or the second most abundant carriers of major elements, like \ce{C2H2} in a \ce{CO}-\ce{CH4} atmosphere (Figure \ref{CHEM_KZZ}). 

Determining the major components and some of the minor components in atmospheres of super Earths and mini Neptunes like GJ~1214~b via spectroscopy may offer a window to their deep atmospheres that are otherwise not detectable, and potentially enable the study of vertical mixing and internal heating on these planets. For a warm exoplanet like GJ~1214~b, we find that for a wide range of composition, and an eddy diffusion coefficient ranging from $10^6\sim10^9$ cm$^2$ s$^{-1}$, the quenching pressure is $10\sim100$ bar. This is true for almost all types of atmospheres.  What determines the abundance of the minor species is the interplay between eddy mixing that determines the quenching pressure, and the internal heat flux that determines the temperature at the quenching pressure. The so called ``minor species'', although low in absolute amounts, may have significant imprints in spectra and therefore be detectable. For example, \ce{C2H2} in a \ce{CH4} dominated atmosphere leads to prominent \ce{C2H2} features in both transmission and thermal emission (Figure \ref{SPEC_FEATURE}). For a very hot super Earth like 55~Cnc~e (equilibrium temperature higher than $\sim2000$ K), however, the atmosphere is very close to thermochemical equilibrium at all relevant pressure levels. For these planets, thermochemical equilibrium calculation would be efficient and adequate for studying the chemical properties of their atmospheres.

\subsection{Atmosphere-Surface Exchange of Super Earths}

A cumulative loss of hydrogen from atmospheres on super Earths would result in the build-up of oxygen, unsaturated hydrocarbons, and other thermochemical and photochemical products. In this study, we neglect the possible material exchange between the atmosphere and the surface of a super Earth. How would active atmosphere-surface exchange affect the atmospheric composition of a super Earth?

First, surface emission of reduced gases (e.g., \ce{H2}, \ce{CH4}, \ce{H2S}) and exposure of unoxidized minerals (e.g., \ce{FeO}) by volcanism can consume atmospheric oxygen and potentially prevent the formation of \ce{O2}-rich atmospheres. Without atmosphere-surface exchange, we predict some super Earths that form in oxygen-rich nebulae will have \ce{O2}-rich atmospheres due to loss of atmospheric hydrogen into the space (Section 3.2). Such \ce{O2}-rich atmospheres may be short-lived due to atmosphere-surface exchange. A Solar-System example reminiscent of this process is Venus. If Venus starts with an ocean and loses its ocean during the runaway greenhouse phase, a scenario supported by the detection of an atmospheric D/H ratio $\sim160$ times higher than the terrestrial value (e.g., Donahue et al. 1982), a massive \ce{O2}-rich atmosphere up to 240 bars must have existed on Venus (Kasting 1997). However, today's Venus atmosphere only has trace amounts of oxygen, which means that that much of oxygen has to be consumed over the history of Venus. Volcanic eruptions can release reduced gases (e.g., \ce{H2}, \ce{CH4}, \ce{H2S}) and expose unoxidized lithosphere (e.g., \ce{FeO}), which can consume atmospheric oxygen. It has been estimated that volcanic eruption rates on Venus has to be a few times higher than the current volcanic eruption rate of Earth to consume the left-over \ce{O2} in 100 million years (Fegley 2008). For a super Earth to prevent the \ce{O2} build up due to hydrogen loss, the level of volcanic activity would need to be at least comparable to Earth's, depending on how fast the hydrogen loss is. 

Second, surface emission of hydrogen and other reduced gases to the evolved atmosphere on a super Earth will be quickly removed by thermochemical reactions that yield energy. For example, in an oxygen-rich atmosphere, emitted hydrogen would react with free oxygen to form water, a process that yields chemical energy. For another example, in an evolved carbon-rich atmosphere, emitted hydrogen would react with unsaturated hydrocarbons to form more saturated forms of carbon, a process that also yields chemical energy. As shown in Figure \ref{WATER_T} and Figure \ref{HYDROCARBON_T}, the scarcity of hydrogen, driven by atmospheric loss, is the controlling factor for the major components in water-rich atmospheres and hydrocarbon-rich atmospheres. Therefore, hydrogen escape to the space could potentially results in a large chemical gradient between the atmosphere and the interior of the planet. Such chemical gradient could be exploited by chemotrophs (organisms that obtain energy through chemical process; see Seager et al. 2012), and if temperature is suitable, lead to potential biosignature gases.

\subsection{Aspects of Further Investigations}

{\it Photochemically Produced Species in the Upper Atmosphere} This paper is mostly concerned with transport-driven disequilibrium chemistry deep in the atmosphere of low-mass exoplanets. However, one should note that photochemical processes can have a large effect on emission and especially transmission spectra, since these spectra are sensitive to the molecular abundances high up in the atmosphere. One example is shown in Figure \ref{CHEM_KZZ}. In this example \ce{C2H2} is produced by photochemical processes above the pressure level of 1 mbar and the photochemical source can increase the \ce{C2H2} mixing ratios to $10^{-2}$. In general, a higher UV flux or a lower eddy diffusivity would favor  the formation of such photochemical derivatives. Further research to quantify the effects of these photochemical products is warranted.

{\it Chemistry of High-Order Hydrocarbons and Haze Formation}
Our current chemical scheme for hydrocarbon chemistry stops at \ce{C2H_x}. For an atmosphere with high $X_{\rm C}/X_{\rm O}$, it is possible to form hydrocarbons with more than 2 carbons, with the chemical paths shown in Yung et al. (1984) for Titan's atmosphere. One might also speculate that for suitable temperatures the photochemically produced hydrocarbons may condense and form haze layers in the atmosphere. In fact, such process has been proposed to occur on some hot jupiters (Zahnle et al. 2009a) and as one of the explanations for the flat transmission spectrum of GJ~1214~b (Morley et al. 2013). However, it is highly complex to model the hydrocarbon chemistry and potential haze formation in planetary atmospheres, because most of the chemical kinetic rates remain unknown.

\section{Conclusion}

We have developed and validated a one-dimensional photochemistry-thermochemistry kinetic-transport model for the exploration of compositions of thick atmospheres on super Earths and mini Neptunes. The unique feature of our photochemistry-thermochemistry model is that our model can treat both \ce{H2}-dominated atmospheres and non-\ce{H2}-dominated atmospheres, and our model is able to compute the main components and the mean molecular mass of an atmosphere based on its elemental abundance. This feature makes our model uniquely suitable for the exploration of super Earth atmospheres, which are expected to have diverse chemical compositions.

Using the photochemistry-thermochemistry model, we have outlined a roadmap to characterize thick atmospheres on super Earths and mini Neptunes. Using the hydrogen abundance ($X_\ce{H}$) and the carbon to oxygen abundance ratio ($X_\ce{C}/X_\ce{O}$) as the primary parameters, we classify thick atmospheres on super Earths and mini Neptunes into hydrogen-rich atmospheres, water-rich atmospheres, oxygen-rich atmospheres, and hydrocarbon-rich atmospheres. We find that when $X_\ce{H}>0.7$, the atmosphere has free hydrogen and chemically behaves like \ce{H2}-dominated atmospheres on gas giants. When $0.3<X_\ce{H}<0.7$, the atmosphere is water-rich for small $X_\ce{C}/X_\ce{O}$ and hydrocarbon-rich for large $X_\ce{C}/X_\ce{O}$. In the middle of these regimes the atmosphere contain \ce{CO} and \ce{H2}, with mixing ratios of \ce{CH4}, \ce{C2H2}, \ce{CO2}, and \ce{H2O} depending on temperature and $X_\ce{C}/X_\ce{O}$ sensitively.

Water-dominated atmospheres, in which most molecules are water vapor, only exist for  $X_\ce{C}/X_\ce{O}<0.2$. We find that in a \ce{H2}-depleted water-dominated atmosphere, most of the trace mixture of carbon has to be in the form of \ce{CO2} rather than \ce{CH4} or \ce{CO}. Therefore a detection of water vapor features together with a confirmation of nonexistence of methane features are sufficient evidence for a water-dominated atmosphere on an exoplanet having similar temperatures as GJ~1214~b. If a water-rich atmosphere continues to lose hydrogen, free oxygen may be left over in the atmosphere to form oxygen-rich atmospheres. 

For hydrocarbon-rich atmospheres, we find that it is the scarcity of hydrogen, which can be a result of preferred loss of light elements or the details of atmospheric outgassing as the planet formed and cooled, that drives the formation of unsaturated hydrocarbon. \ce{C2H2} and \ce{C2H4} can be the dominant forms of carbon in some cases, and they are the hallmark molecules for carbon-rich atmospheres on super Earths. Therefore they should be considered among ``standard'' building blocks for atmospheres on super Earths. Also for hydrocarbon-rich atmospheres we find that \ce{H2O} should be scarce and most of oxygen should be in the form of \ce{CO}. 

In terms of observational characterization of atmospheres on super Earths, our photochemistry-thermochemistry models can eliminate a significant part of molecular composition parameter space. We show that the carbon to oxygen elemental abundance ratio is the key parameter that defines thick atmospheres on super Earths, and this parameter can be constrained by detecting spectral features of hallmark molecules that indicate the atmospheric scenarios (see the cookbook for observers in Section 3.1). 

\begin{figure}[h]
\begin{center}
 \includegraphics[width=0.7\textwidth]{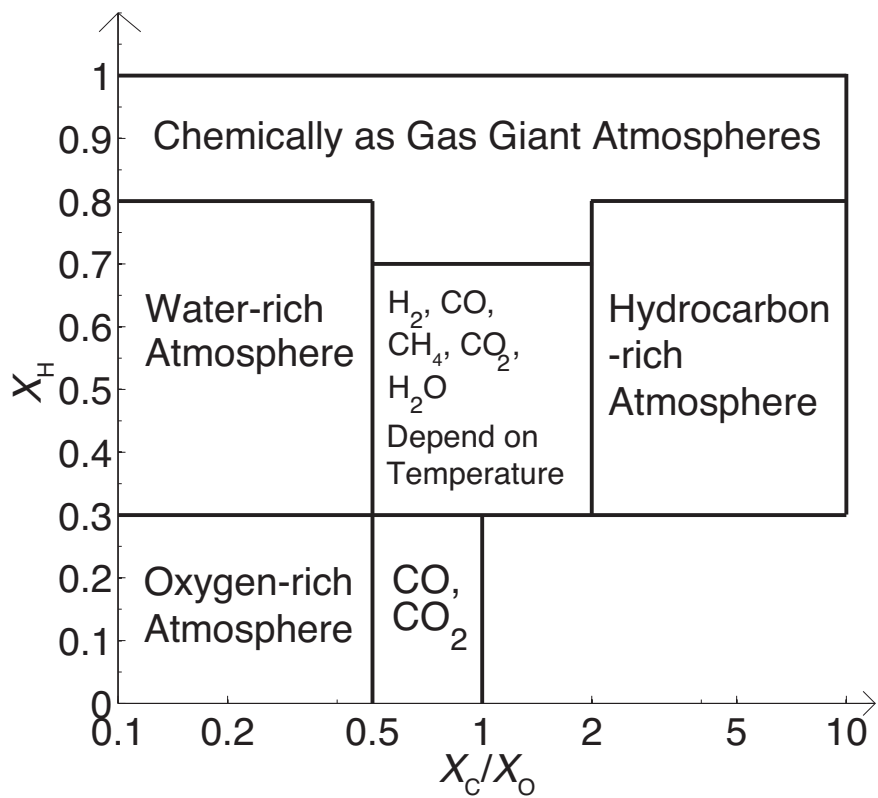}
 \caption{
 Chemical classification of thick atmospheres on super Earths and mini Neptunes. The classification is summarized based on extensive numerical exploration for exoplanets with equilibrium temperatures ranging from 500 to 2000 K, on a 2-dimensional grid that explores the hydrogen abundance and the carbon to oxygen abundance ratio in the atmosphere. This figure shows the theoretical range of atmospheric chemical composition; reality may be more restrictive. The main point is that, when \ce{H2} is no longer the dominant component in the atmosphere, water-rich atmospheres, hydrocarbon-rich atmospheres, and oxygen-rich atmospheres emerge, depending on the hydrogen abundance and the carbon to oxygen ratio. In the middle of these regimes is where \ce{H2}, \ce{CO}, \ce{CH4}, \ce{CO2}, and \ce{H2O} can coexist, and their relative abundances are determined by the temperature and the elemental abundance of the atmosphere.
  }
 \label{Zoo}
  \end{center}
\end{figure}

Our classification of thick atmospheres on super Earths and mini Neptunes is closely related to the formation and evolution of low-mass exoplanets. All parts of the diagram in Figure \ref{Zoo} may not be equally likely, and we will rely on the formation and evolution models for prior likelihood of different parts of the diagram. Populating the classification diagram by future observations in synergy with planet formation models for a number of super Earths and mini Neptunes will greatly improve our understanding of the formation and evolution of low-mass exoplanets.
%In particular, for warm super Earths and mini Neptunes like GJ~1214~b, we find \ce{C2H2} and \ce{C2H4} are the diagnostic molecules for carbon-rich atmospheres; \ce{CH4} should not be expected in a \ce{H2}-depleted water-dominated atmosphere; and \ce{H2O} should be scarce in carbon-rich atmospheres. For hot super Earths like 55~Cnc~e, we find \ce{CH4} should not be expected in any cases; and \ce{CO} and \ce{C2H2} are the diagnostic molecules for carbon-rich atmospheres. 
%The ranges of molecular compositions presented in this paper that contain hallmark molecules will allow observational characterization of different atmosphere scenarios on super Earths and mini Neptunes.

\acknowledgments

Some of the computation presented in this paper is performed at California Institute of Technology, and RH is grateful to Yuk L. Yung for hospitality. We thank the anonymous reviewer for the insightful comments that helped improve the paper. We thank C. Visscher for providing thermochemistry data of several species. This work has made use of the MUSCLES M-dwarf UV radiation field database. This work was supported by NASA Headquarters under the NASA Earth and Space Science Fellowship Program - Grant ``NNX11AP47H''.

\clearpage

\begin{table}[htdp]
\caption{The critical temperature and critical pressure for common building blocks for planetary thick atmospheres.}
\begin{center}
\begin{tabular}{ccc}
\hline\hline
Substance & Critical Temperature (K) & Critical Pressure (bar) \\
\hline
\ce{H2} & $33.2\pm0.2$ & $13.00\pm0.01$ \\
\ce{H2O} & $647\pm2$ & $220.64\pm0.05$ \\
\ce{CH4} & $190.6\pm0.3$ & $46.1\pm0.3$ \\
\ce{CO} & $134.5\pm0.4$ & $35.0\pm0.3$ \\
\ce{CO2} & $304.18\pm0.04$ & $73.80\pm0.15$ \\
\ce{O2} & $154.58\pm0.0015$ & $50.43\pm0.005$ \\
\ce{C2H2} & $308.3\pm0.1$ & $61.4\pm0.1$ \\
\ce{C2H4} & $282.5\pm0.5$ & $50.6\pm0.5$ \\
\ce{C2H6} & $305.3\pm0.3$ & $49\pm1$ \\
\ce{N2} & $126.19\pm0.01$ & $33.978\pm0.007$ \\
\ce{NH3}& $405.4\pm0.1$ & $113.00\pm0.05$ \\
\hline\hline
\end{tabular}
\tablecomments{The data are compiled from the NIST Chemistry WebBook (http://webbook.nist.gov/chemistry/).}
\end{center}
\label{critical}
\end{table}

\clearpage

\begin{landscape}

\begin{scriptsize}

\begin{table*}[htdp]
\begin{scriptsize}
\caption{Main components of thick atmospheres on super Earths and mini Neptunes as a function of the abundances of hydrogen, oxygen, and carbon.}
\begin{center}
\begin{tabular}{llll}
\hline\hline
Parameter Regime & Main Components & Mixing Ratio Formulae & Corresponding Classification \\
\hline
$X_\ce{H}\sim1$, $X_\ce{O}\ll1$, $X_\ce{C}\ll1$ & \ce{H2} & $X_\ce{H2}\sim1$ & Hydrogen-dominated atmospheres\\
$X_\ce{H}\ll1$, $X_\ce{O}\sim1$, $X_\ce{C}\ll1$ & \ce{O2} & $X_\ce{O2}\sim1$ & Oxygen-rich atmospheres \\
$X_\ce{H}\sim1$, $X_\ce{O}\sim1$, $X_\ce{C}\ll1$, $X_\ce{H}\geq 2X_\ce{O}$ & \ce{H2}, \ce{H2O} & $X_\ce{H2}=1-\frac{2X_\ce{O}}{X_\ce{H}}$, $X_\ce{H2O} = \frac{2X_\ce{O}}{X_\ce{H}}$ & Water-dominated atmospheres\\
$X_\ce{H}\sim1$, $X_\ce{O}\sim1$, $X_\ce{C}\ll1$, $X_\ce{H}<2X_\ce{O}$   & \ce{H2O}, \ce{O2} & $X_\ce{H2O}=\frac{2X_\ce{H}}{X_\ce{H}+2X_\ce{O}}$, $X_\ce{O2}=\frac{2X_\ce{O}-X_\ce{H}}{X_\ce{H}+2X_\ce{O}}$ & Water-dominated atmospheres \\
$X_\ce{H}\sim1$, $X_\ce{O}\ll1$, $X_\ce{C}\sim1$      & \ce{H2}, \ce{CH4}, \ce{C2H2}, \ce{C2H4}, \ce{C2H} & Temperature dependent  & Hydrocarbon-rich atmospheres\\
$X_\ce{H}\ll1$, $X_\ce{O}\sim1$, $X_\ce{C}\sim1$, $X_\ce{C}\leq X_\ce{O}\leq 2X_\ce{C}$ & \ce{CO}, \ce{CO2} & $X_\ce{CO}=2-\frac{X_\ce{O}}{X_\ce{C}}$, $X_\ce{CO2} = \frac{X_\ce{O}}{X_\ce{C}}-1$ & \ce{CO}-\ce{CO2} atmospheres\\
$X_\ce{H}\ll1$, $X_\ce{O}\sim1$, $X_\ce{C}\sim1$, $X_\ce{O}>2X_\ce{C}$ & \ce{CO2}, \ce{O2} & $X_\ce{CO2} = \frac{2X_\ce{C}}{X_\ce{O}}$, $X_\ce{O2} = 1-\frac{2X_\ce{C}}{X_\ce{O}}$ & Oxygen-rich atmospheres\\
$X_\ce{H}\sim1$, $X_\ce{O}\sim1$, $X_\ce{C}\sim1$ & \ce{CO}, \ce{H2}, \ce{H2O}, \ce{CH4}, \ce{CO2} & Temperature dependent & \\
\hline
\hline
\end{tabular}
\tablecomments{The table lists all plausible combinations of H, O, and C, in the orders of one-element dominance, two-element dominance, and three-element dominance, except for the case of $X_\ce{H}\ll1$, $X_\ce{O}\ll1$, $X_\ce{C}\sim1$ as elemental carbon is not in the gaseous phase for the temperatures of interest. The main components are stable molecules made of H, O, and C, and the table is valid to the extent that the molecular forms are thermochemically favored over elemental forms and water vapor does not condense in the atmosphere.}
\end{center}
\label{Para}
\end{scriptsize}
\end{table*}

\end{scriptsize}

\end{landscape}

\clearpage

\clearpage

\clearpage

\clearpage

\clearpage

\clearpage

\clearpage

\clearpage

\clearpage

\clearpage

\clearpage

\clearpage

%\clearpage
%
%\begin{figure}[h]
%\begin{center}
% \includegraphics[width=0.7\textwidth]{GJ1214B_hydrocarbonrich.eps}
% \caption{
% Molecular composition of hydrocarbon-rich atmospheres on GJ~1214~b like super Earths/mini Neptunes, with same format as Figure \ref{CHEM_KZZ}. For $X_\ce{C}/X_\ce{O}>1$, \ce{CO}, \ce{CH4}, \ce{C2H2}, and \ce{C2H4} all have mixing ratio in the order of 0.1 at the pressure level of $0.1\sim10$ mbar; and the exact compositions are sensitive to the eddy diffusion coefficient.
%  }
% \label{HYDROCARBON}
%  \end{center}
%\end{figure}

\clearpage

\clearpage

\clearpage

\clearpage

%\clearpage
%
%\begin{figure*}[h]
%\begin{center}
% \includegraphics[width=0.95\textwidth]{REFL_DM.eps}
% \caption{
% Modeled reflection spectra of thick atmospheres on a GJ~1214~b sized exoplanet at 1-AU orbit of a Sun-like star based on photochemistry-thermochemistry simulations. The presented spectra are for both \ce{H2}-dominated ($X_\ce{H}=0.99$) and non-\ce{H2}-dominated ($X_\ce{H}=0.5$) atmospheres with a variety of carbon to oxygen ratios ranging from oxygen-rich to carbon-rich, and for an eddy diffusivity of $10^9$ cm$^2$ s$^{-1}$. A cloud deck is expected at $\sim10$ mbar for these scenarios, composed of either water droplets or photochemical haze. We assume that the cloud deck has a uniform albedo of 0.3 for this calculation, and therefore potential cloud-related spectral features are not shown here.
% The reflection spectra are expected to be dominated by methane features.
% }
% \label{SPEC_DM}
%  \end{center}
%\end{figure*}

\clearpage

\clearpage

\end{document}